\documentclass[onecolumn,notitlepage]{revtex4-1}

\usepackage{graphicx}
\usepackage{amsmath}
\usepackage{amssymb}

\newcommand{\onehalf}{\frac{1}{2}}
\newcommand{\be}{\begin{equation}}
\newcommand{\ee}{\end{equation}}
\newcommand{\bea}{\begin{eqnarray}}
\newcommand{\eea}{\end{eqnarray}}
\newcommand{\nn}{\nonumber}

\newcommand{\RR}{\mathbf{r}}
\newcommand{\rr}{\mathbf{r}}

\newcommand{\dr}{d\mathbf{r}}

\newcommand{\kk}{\mathbf{k}}

\newcommand{\rhon}{{n({\mathbf r})}}

\newcommand{\rhorOm}{{\rho\left({\mathbf r},{\mathbf u}\right)}}
\newcommand{\rhorom}{{\rho\left({\mathbf r},\omega\right)}}

\newcommand{\alpharom}{{\alpha\left({\mathbf r},\omega\right)}}

\newcommand{\sigrom}{{\sigma \left({\mathbf r},\omega\right)}}
\newcommand{\Om}{\mathbf{u}}
\newcommand{\Frho}{{\cal F}[\rho]}
\newcommand{\F}{{\cal F}}
\newcommand{\Rhat}{\hat{{\mathbf r}}}
\newcommand{\Pol}{\mathbf{P}(\mathbf {r})}

\newcommand{\MU}{\boldsymbol{\mu}}

\newcommand{\PP}{\mathbf{P}}
\newcommand{\EE}{\mathbf{E}}

\newcommand{\Pon}{\frac{P({\mathbf r})}{\mu \, n({\mathbf r})}}
\newcommand{\Ld}{{\cal L}}
\newcommand{\Lm}{{\cal L}^{-1}}

\begin{document}

\title{Molecular density functional theory of water including density-polarization coupling}

\author{Guillaume Jeanmairet}
\affiliation{Max Planck Institute for Solid State Research, Heisenbergstrasse 1, Stuttgart 70569, Germany}

\author{Nicolas Levy}
\affiliation{\'Ecole Normale Sup\'erieure, PSL Research University, D\'epartement de Chimie, Sorbonne Universit\'es -- UPMC Universit\'e Paris 06, CNRS UMR 8640 PASTEUR, 24 rue Lhomond, 75005 Paris, France}

\author{Maximilien Levesque}
\email{maximilien.levesque@ens.fr}
\affiliation{\'Ecole Normale Sup\'erieure, PSL Research University, D\'epartement de Chimie, Sorbonne Universit\'es -- UPMC Universit\'e Paris 06, CNRS UMR 8640 PASTEUR, 24 rue Lhomond, 75005 Paris, France}

\author{Daniel Borgis}
\affiliation{\'Ecole Normale Sup\'erieure, PSL Research University, D\'epartement de Chimie, Sorbonne Universit\'es -- UPMC Universit\'e Paris 06, CNRS UMR 8640 PASTEUR, 24 rue Lhomond, 75005 Paris, France}
\affiliation{Maison de la Simulation, USR 3441, CEA-CNRS-INRIA- Universit\'e Paris-Sud - Universit\'e de Versailles, 91191 Gif-sur-Yvette, France}

\begin{abstract}
We present a three-dimensional molecular density functional theory (MDFT) of water derived from first-principles that relies on the particle's density and multipolar polarization density and includes the density-polarization coupling. This brings two main benefits: ($i$) a scalar density and a vectorial multipolar polarization density fields are much more tractable and give more physical insight than the full position and orientation densities, and ($ii$) it includes the full density-polarization coupling of water, that is known to be non-vanishing but has never been taken into account. Furthermore, the theory requires only the partial charge distribution of a water molecule and three measurable bulk properties, namely the structure factor and the Fourier components of the longitudinal and transverse dielectric susceptibilities.
\end{abstract}

\maketitle

\section{Introduction}

Numerical methods based on liquid state theories\cite{hansen, gray-gubbins-vol1} are now at the heart of many physical chemistry or chemical engineering applications\cite{gray-gubbins-vol2,neimark06,neimark11,wu06}: they can predict both the structure and thermodynamic properties of molecular fluids at a much lower computational cost than molecular dynamics or Monte Carlo simulations.
Those methods include integral equation theory in the interaction-site\cite{Chandler-RISM,hirata-rossky81,hirata-pettitt-rossky82,reddy03,pettitt07,pettitt08}
or molecular\cite{blum72a,blum72b,patey77,fries-patey85,richardi98,richardi99,belloni12,belloni14} picture, classical density functional theory (DFT)\cite{evans79,evans92,wu07}, or classical fields theory\cite{chandler93,lum99,tenwolde01,coalson96}.

A current trend is their challenging implementation in three dimensions. This would unlock great technological applications like the description of molecular liquids, solutions, and mixtures in complex environments such as solid interfaces or bio-molecular media at the atomic scale. Tremendous efforts have been ongoing in this direction, especially from the 3D-RISM\cite{Beglov-Roux97,kovalenko-hirata98,red-book,yoshida09,kloss08-jcp,kloss08-jpcb}, lattice field\cite{azuara06,azuara08} or Gaussian field\cite{tenwolde01,varilly11} theories.

Classical density functional theory (DFT) has been much less developed for chemical applications. Its theoretical principles can be found in the seminal papers by Evans\cite{evans79} and subsequent reviews\cite{evans79,henderson_fundamentals_1992,evans_density_2009,lowen_density_2002}. Nevertheless, since the advent of quasi-exact DFT for hard spheres and their mixtures,  generated much attention, developments and success for atom-like fluids, in bulk and confined systems.\cite{rosenfeld89,kierlik-rosinberg90,kierlik-rosinberg91,roth02,yu_structures_2002,roth-review10,wu07,wu09}
Much less applications exist for molecular fluids, for which solvent orientations should be considered. The DFTs proposed so far to mimic molecular solvents are generally limited to generic dipolar solvents\cite{telodagama91,dietrich92} or mixtures of dipolar solvents and ions\cite{biben98,oleksy_microscopic_2009,oleksy_wetting_2010,oleksy_wetting_2011}. Even if the solvent model is simplified, it is there already a great step forward compared to primitive continuum models of solvation: They are considered "civilized".\cite{oleksy_wetting_2010}

For describing the solvation of three-dimensional molecular object in arbitrary solvents, we have introduced a molecular density functional theory (MDFT) approach to solvation\cite{ramirez02,ramirez05-CP,ramirez05,gendre09,zhao11,borgis12,levesque12_1,levesque12_2,jeanmairet_molecular_2013-1,jeanmairet_molecular_2013,jeanmairet_hydration_2014,sergiievskyi_fast_2014}.  In the general case, it relies on the definition of a free-energy functional depending on the full six-dimensional position and orientation solvent density. In the so-called homogeneous reference fluid (HRF) approximation, the (unknown) excess free energy can be inferred from the angular-dependent direct correlation function of the bulk solvent. This last quantity can be predetermined from molecular simulations of the pure solvent. Compared to reference molecular dynamics calculations, such approximation was shown to be accurate for polar, non-hydrogen bonded fluids \cite{ramirez02,gendre09,zhao11,borgis12,levesque12_2}. Nevertheless, it requires corrections for water\cite{zhao11,zhao-wu11,zhao-wu11-correction,levesque12_1}. Note that a RISM-based DFT approach of molecular solvation has been developed recently\cite{liu_site_2013}.

Recently, we have introduced a molecular density functional theory (MDFT) for water models of the kind of SPC, SPC/E or TIPnP.~\cite{jeanmairet_molecular_2013} These common models of water have in common that they interact with each other with a single Lennard-Jones site and distributed partial charges. For these cases, we showed how to write a functional of the water density $n(\rr)$ and site-distributed polarization density $\PP(\rr)$. This functional only requires simpler physical quantities than the full position and orientation-dependent direct correlation function. This functional can be motivated from first-principles, or may be considered as a multipolar generalization of the generic dipolar fluid free-energy functional $\F[n(\rr),\PP(\rr)]$ introduced in Refs.~\cite{ramirez02,ramirez05}. These are the routes which are detailed here, \emph{and generalized}.

\section{General molecular functional}

\label{sec:general_functional}

In the general case of solvation of molecular entities in an arbitrary molecular solvent, the solvent molecules are considered as rigid entities with position $\rr$ and orientation $\omega$ (in terms of three Euler angles $\theta, \phi, \psi$ in a fixed frame). The system is characterized by the position and orientation density $\rhorom$ and by the density free-energy functional
\begin{equation}
\Frho = \F_{id}[\rho] + \F_{ext}[\rho] + \F_{exc}[\rho],
\label{eq:exact-functional}
\end{equation}
with the following expressions of the ideal, external, and excess terms:
\begin{eqnarray}
\hspace{-1cm} \F_{id}[\rho]  \!\!&=&\!\!
k_BT \int d\rr d\omega \left[ \rhorom  {\rm ln}\left(\frac{8\pi^2 \rhorom}{n_0}\right )
 - \rhorom+\frac{n_0}{8\pi^2} \right] , \label{eq:ideal} \\
\nn \\
\F_{ext}[\rho] &=& \int d\rr d\omega\,
V_{ext}(\rr,\omega) \rhorom
,
\label{eq:externo} \\
\nn \\
 \F_{exc}[\rho]\!\!&=&\!\! -\onehalf k_BT\int\!\!\int d\rr d\rr' d\omega  d\omega'  \Delta\rhorom
 \, c(|\rr - \rr'|,\omega,\omega') \,
 \Delta \rho(\rr',\omega') \, + \F_B[\rho]
\label{eq:excess}
\end{eqnarray}
where $n_0$ is the solvent bulk density (for instance $n_0=0.033$ \AA$^{-3}$  for water at 1~atm and 300~K), and $\Delta\rhorom =\rhorom - n_0/8\pi^2$. In the last equation, the first term represents  the homogeneous reference fluid (HRF) approximation (equivalent to a  solute-solvent HNC approximation) where the excess free-energy density is written in terms of the angular-dependent direct correlation of the {\em pure} solvent. The second term (the so-called bridge term)  represents the unknown correction to the exact functional, and  can be expressed as of a systematic expansion of the solvent-solvent correlations in terms of three-body, $\ldots$ n-body  direct correlation functions. 
The second-order direct correlation function $c$ is related to the angular-dependent pair distribution function $h$ by the homogeneous-fluid Ornstein-Zernike (OZ) equation that reads in $k$-space
\be
h(\kk,\omega,\omega^\prime) = c(\kk,\omega,\omega^\prime) + \rho_0 \int d\omega^{\prime\prime} \, c(\kk,\omega,\omega^{\prime\prime}) h(\kk,\omega^{\prime\prime},\omega^{\prime})  \label{eq:angular_OZ}
\ee
The minimization of the functional of Eqs.\ref{eq:ideal}-\ref{eq:excess} to get the equilibrium density and solvation free energy for a given three-dimensional potential, as well as the resolution of the OZ equation for $c$ knowing $h$, was implemented numerically using a three-dimensional grid for positions and an angular grid on each $\rr$ or $\kk$-point for the orientations. This methodology is described in Refs~\cite{gendre09,zhao11,borgis12}. The functional minimization is performed with respect to $\rhorom$ in $\rr$-space, and the bottleneck is indeed the computation of the excess free energy, which has to be  done in $\kk$-space using fast Fourier transforms (FFT) to avoid the numerical cost of the double integration over $\rr$. Up to now the angular dependence is handled by direct integration over the angular grid, which implies a double integration in Eq.~\ref{eq:excess}. We are presently implementing an angular Fourier transforms algorithm to improve the efficiency. This requires a systematic expansion of the direct correlation function into a rotational invariants basis set
 \bea
c(\rr_{12},\omega_1,\omega_2) &= & \sum_{mnl, \mu \nu} c^{lmn}_{\mu \nu}(r_{12})\,\Phi^{mnl}_{\mu \nu}(\omega_1,\omega_2),
\label{eq:c-functions}
\eea
and a similar expansion for $h(\rr_{12},\omega_1,\omega_2)$ to solve the OZ equation\cite{blum72a,blum72b,patey77,fries-patey85,richardi98,richardi99,belloni12,belloni14}. 

From  this point we depart from the general case described above and turn to the search of  functionals applying to a special class of solvent models, which can be termed as simple point charge models, involving  a single Lennard-Jones site embedding a charge distribution, which can be described either by a set of distributed point charges, or by a series of multipoles located at the center. This class of models is indeed  important since it includes the prototypical Stockmayer model as well as the most widely used water models such as SPC, SPCE, TIP3P, TIP4P, $\ldots$

\section{Functional at dipolar order}

\label{sec:dipolar}
\subsection{Dipolar functional including density-polarization coupling}
To progress towards the definition of a general functional for SPC-like  models, we next  suppose  that the systematic expansion of $c(\rr_{12},\omega_1,\omega_2)$ into rotational invariant can be limited to dipolar order, that is  to $m, n  \le 1$ and $\mu, \nu = 0$. The latter condition apply to dipolar symmetry, or $C_{2v}$ symmetry as for water, where $\mu, \nu$ should be even. Omitting those
two indices for clarity, the relevant rotational invariants read 
\begin{eqnarray}
\Phi^{0 0 0} & =  & 1 \nn \\
\Phi^{1 0    1} & =  &  \Om_1 \cdot \Rhat_{12} \nn \\
\Phi^{0  1 1} & =  & -  \Om_2 \cdot \Rhat_{12} \nn \\
\Phi^{1 1 0}&=&\Om_1 \cdot \Om_2 \nn
\\ \Phi^{1 1 2}&=&3\,(
\Om_1\cdot\Rhat_{12})\,(
\Om_2\cdot\Rhat_{12})-  \Om_1 \cdot \Om_2 . 
\end{eqnarray}
They are defined in terms of the intermolecular unit vector, $\Rhat_{12} = \mathbf r_{12}/r_{12}$, and the orientation unit vector for each molecular  dipole,  $\Om_i$. They thus depend on  the angles 
$\theta_i,\phi_i$ but are independent of $\psi_i$.  
 We also suppose momentarily that the  bridge term can be neglected, $\F_B[\rho] = 0$.
 
Injecting the corresponding c-function into the general density functional expression, Eqs~\ref{eq:exact-functional} to \ref{eq:excess}, leads to  the following functional depending on $\rhorOm=\int d\psi \rho(\rr,\Om,\Psi)$ only
\bea
\Frho &= & k_BT \int d\rr d\Om \left[ \rhorOm{\rm ln}\left(\frac{4\pi \rhorOm}{n_0}\right ) - \rhorOm + \frac{n_0}{4\pi} \right] \nn \\
&+&  \int d\rr d\Om \, V_{eff}(\rr,\Om) \rhorOm  + \F_{exc}[n,\PP],
\eea
where $V_{eff}(\rr,\Om)$ is an effective $\Om$-dependent external potential defined as
  \begin{equation}
V_{eff}(\rr,\Om)=- k_BT \ln \left[ \frac{1}{2\pi} \int d\Psi \exp \left( -\beta V_{ext}(\rr,\Om,\Psi) \right) \right],
\end{equation}
and where the excess free-energy turns out to depend on $\rhorOm$ through the number density
and the polarization density,
\bea
\rhon &=&\int d\Om \, \rho(\RR,\Om) \\
\Pol &=&\ \mu \int d\Om\, \Om \, \rho(\RR,\Om), \label{eq:Pol_def}
\eea
where $\mu$ is the amplitude of the dipole of the solvent. More specifically 
\bea
\beta \F_{exc}[n,\PP] &=& - \onehalf  \int d\rr_1 d\rr_2  \, c^{000}(r_{12}) \, \Delta n(\rr_1) \Delta n(\rr_2)\nn \\
 &-& \frac{1}{\mu}  \int d\rr_1 d\rr_2 \, c^{101}(r_{12}) \, (\mathbf{P}(\rr_1) \cdot \Rhat_{12}) \, \Delta n(\rr_2) \nn \\ 
&- &\frac{1}{2 \mu^2}  \int d\rr_1 d\rr_2 \, c^{110}(r_{12}) \mathbf{P}(\rr_1) \cdot  \mathbf{P}(\rr_2)  \label{eq:Fdip} \\
& - & \frac{1}{2 \mu^2} \int d\rr_1 d\rr_2    \, c^{112}(r_{12}) \, \left[ 3 \, (\mathbf{P}(\rr_1) \cdot\Rhat_{12}) \,
(\mathbf{P}(\rr_2)\cdot\Rhat_{12})-  \mathbf{P}(\rr_1) \cdot \mathbf{P}(\rr_2) \right]\nn,
\eea
with $\Delta n(\rr) = n(\rr) - n_0$.  We further introduce the longitudinal and transverse polarization in k-space
\bea
\PP(\kk) &= & \PP_L(\kk) + \PP_T(\kk) \\
\PP_L(\kk) &= &(\PP(\kk) \cdot \hat{\kk}) \, \hat{\kk} =P_L(\kk) \hat{\kk},
\eea
with $\hat{\kk}  = \kk/k$, such that the polarization charge density is defined by $\rho_P(\rr) = - \nabla \cdot \Pol$ (as in continuum electrostatics), i.e.,
\be
\rho_P(\kk) = -i \kk \cdot \PP(\kk) = -ik \, P_L(\kk).
\ee
The  excess free energy of eq.\ref{eq:Fdip} can thus be written in k-space
\bea
\beta \F_{exc} &=&  - \frac{1}{2} \int d\kk  \, c^{000}(k) \, \Delta  n(\kk) \Delta n(-\kk) - \frac{i}{\mu} \int d\kk \, c^{101}(k) \,  (\PP(\kk) \cdot \hat{\kk}) \, n(-\kk) \nn \\
&-&\frac{1}{2 \mu^2}  \int d\kk \, c^{110}(k) \, \PP(\kk) \cdot \PP(-\kk) \nn \\
&-&  \frac{1}{2 \mu^2} \int d\kk \, c^{112}(k) \, \left[3(\PP(\kk) \cdot \hat{\kk})(\PP(-\kk) \cdot \hat{\kk}) - \PP(\kk) \cdot \PP(-\kk) \right],
\eea
where we have introduced the Fourier-Bessel transform (also known as Hankel transform) of the spherically symmetric, radial projections.
We note that projections $101$ and $011$ are purely imaginary in k-space and noted as $ic^{101}(k) = - i c^{011}(k)$. Owing to
\bea
\PP(\kk) \cdot \PP(-\kk) &=& \PP_L(\kk) \cdot \PP_L(-\kk) + \PP_T(\kk) \cdot \PP_T(-\kk) \\
\PP_L(\kk) \cdot \PP_L(-\kk) & = & (\PP(\kk) \cdot \hat{\kk})(\PP(-\kk) \cdot \hat{\kk}) 
\eea
we find easily
\bea
\F_{exc}[n,\PP] &=& - \frac{1}{2}  \int d\kk \, c_{n}(k) \, \Delta n(\kk) \Delta n(-\kk)  - \frac{i}{\mu}  \int d\kk \, c_{nP}(k) \, P_L(\kk) \Delta n(-\kk) \nn \\
& -& \frac{1}{2 \mu^2}  \left[  \int d\kk \, c_L(k) \,  \PP_L(\kk) \cdot \PP_L(-\kk) + \int d\kk \, c_T(k) \, \PP_T(\kk) \cdot \PP_T(-\kk)  \right] \label{eq:Fexc_P_k-space_with_C}
\eea
with $c_n(k) = c^{000}(k), \, c_{nP}(k) = c^{101}(k)$  and the usual combinations\cite{gendre-these,raineri92,raineri93}:
\bea
 c_L(k)& =& c^{110}(k) - c^{112}(k) \\
 c_T(k)& =& c^{110}(k) + 2 c^{112}(k).
 \label{eq:c+-}
 \eea
 Defining the particle density susceptibility by
 \be
\chi_n(k) = 1 - n_0 h^{000}(k), \label{eq:chi_n_dipolar}
\ee 
and, according to Refs~\cite{raineri92,raineri93}, the longitudinal and transverse susceptibilities, or the longitudinal and transverse dielectric constant
 by
 \bea
\chi_L =  1 - \frac{1}{\epsilon_L(k)} &=& 3y \left( 1 + \frac{n_0}{3} \, h_L(k) \right)  \label{eq:chi_L_dipolar} \\
 4 \pi \chi_T = \epsilon_T(k) - 1 & = & 3y \left( 1 + \frac{n_0}{3} \, h_T(k) \right) \label{eq:chi_T_dipolar}
 \eea
with $y = \beta \mu^2 n_0/9\epsilon_0 = (\mu/3\mu_0)^2$,  a reference dipole being defined by $\mu_0 = (\epsilon_0/\beta n_0)^{1/2}$. $h_L(k)$ and $h_T(k)$ are defined as in eq.~\ref{eq:c+-}, and  the coupled density-polarization susceptibility is equal to
\be
\chi_{nL}(k) =  \sqrt{y} \, h^{101}(k).
\ee
Using e.g. the $\chi$-transform procedure\cite{blum72b,belloni14}, the OZ equation \ref{eq:angular_OZ} can be inverted to 
give
 \bea
1 -  n_0 c_n(k)& = & \chi_n^{-1}(k) = \frac{\chi_L(k)}{\chi_n(k) \chi_L(k) - \chi_{nL}(k)^2} \nn \\
 1 - \frac{n_0}{3} c_L(k) &=&  3y \, \chi_L^{-1}(k) =  \frac{3y \, \chi_n(k)}{\chi_n(k) \chi_L(k) - \chi_{nL}(k)^2} \label{eq:inverse_chi}\\
 n_0 c^{101}(k) &= & - 3\sqrt{y} \, \chi_{nL}^{-1}(k) = - \frac{3\sqrt{y} \, \chi_{nL}}{\chi_n(k) \chi_L(k) - \chi_{nL}(k)^2} \nn \\
 1 - \frac{n_0}{3} c_T(k) &= &  \frac{3y}{4 \pi \chi_T(k)} 
 \eea
 The excess free energy including all the density and polarization terms can thus be written  in k-space
 \bea
 \F_{exc}[n,\PP]&=&  -\frac{k_BT}{2 n_0}  \int d\kk \Delta n(\kk) \, \Delta n(-\kk) -\frac{3k_BT}{2n_0\mu^2}  \int d\kk \, \PP(\kk) \cdot \PP(-\kk) \nn \\
 &-& \frac{k_BT}{2 n_0}  \int d\kk \, \chi_n^{-1}(k) \, \Delta n(\kk) \, \Delta n(-\kk) + \frac{i  k_BT}{n_0 \mu_0}  \int d\kk \, \chi_{nL}^{-1}(k) \, P_L(\kk) \Delta n(-\kk) \nn \\
&+& \frac{1}{8\pi \epsilon_0}  \int d\kk \, 4 \pi \chi_L^{-1}(k) \, \PP_L(\kk) \cdot \PP_L(-\kk) \nn\\
&+& \frac{1}{8\pi \epsilon_0}  \int d\kk \,  \chi_T^{-1}(k) \, \PP_T(\kk) \cdot \PP_T(-\kk)   \label{eq:Fexc_P_k-space}
\eea
which, using the polarization charge density, can also be written as
\bea
\F_{exc}[n,\PP] &=&  -\frac{k_BT}{2 n_0}  \int d\kk \Delta n(\kk) \, \Delta n(-\kk) -\frac{3k_BT}{2 n_0 \mu^2}  \int d\kk \, \PP(\kk) \cdot \PP(-\kk)  \nn \\
 &+& \frac{k_BT}{2 n_0}  \int d\kk \, \chi_n^{-1}(k) \, \Delta n(\kk) \, \Delta n(-\kk) + \frac{k_BT}{n_0 \mu_0}  \int d\kk \, \frac{1}{k}\chi_{nL}^{-1}(k) \, \rho_P(\kk) \Delta n(-\kk) \nn \\
&+& \frac{1}{8\pi \epsilon_0}  \int d\kk \, \frac{4 \pi}{k^2} \chi_L^{-1}(k) \, \rho_P(\kk) \rho_P(-\kk) \nn\\
&+& \frac{1}{8\pi \epsilon_0}  \int d\kk \,  \chi_T(k)^{-1} \, \PP_T(\kk) \cdot \PP_T(-\kk) \label{eq:Fexc_rho_k-space} 
\eea
One recognizes in the charge density-charge density interaction the usual Coulombic interaction $4\pi/k^2$ damped by the inverse susceptibility $\chi_L^{-1}(k)$; this converts in r-space to an effective Coulombic interaction $1/\left(r S(r)\right)$, defined as the inverse Fourier transform of  $4 \pi \chi_L^{-1}(k)/k^2$.

\subsection{The Stockmayer solvent}

\begin{figure}
	\includegraphics[width=8cm]{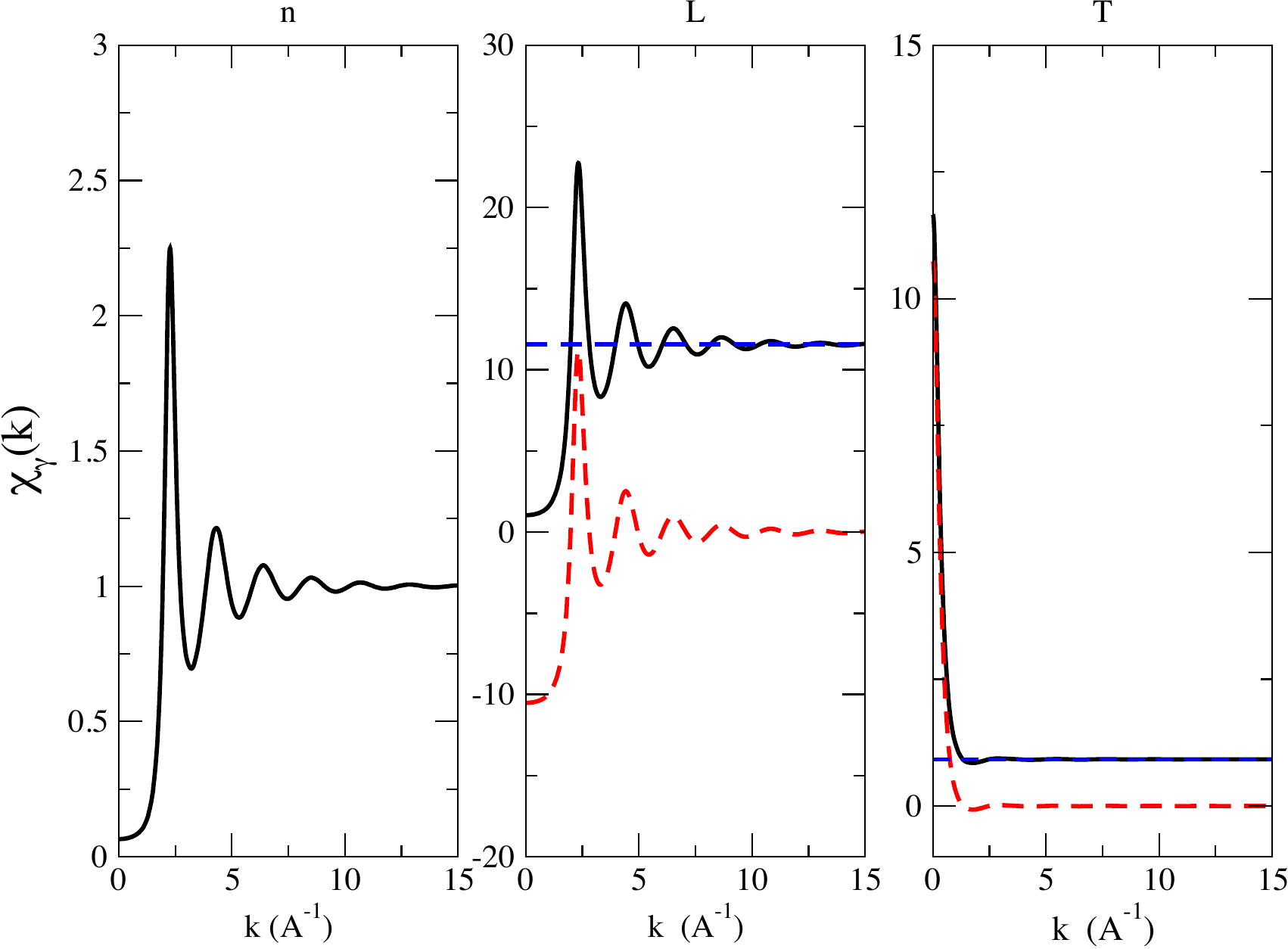}
	\caption{ \label{fig:chi_nLT_stockmayer}
Susceptibilities $\chi_\gamma(k)$ with $\gamma = n, L, T$  for  the Stockmayer model described in the text, computed from the total correlation functions as in eqs~\ref{eq:chi_n_dipolar}. The latter functions were obtained from  MD simulations.}
\end{figure}

\begin{figure}
	\includegraphics[width=8cm]{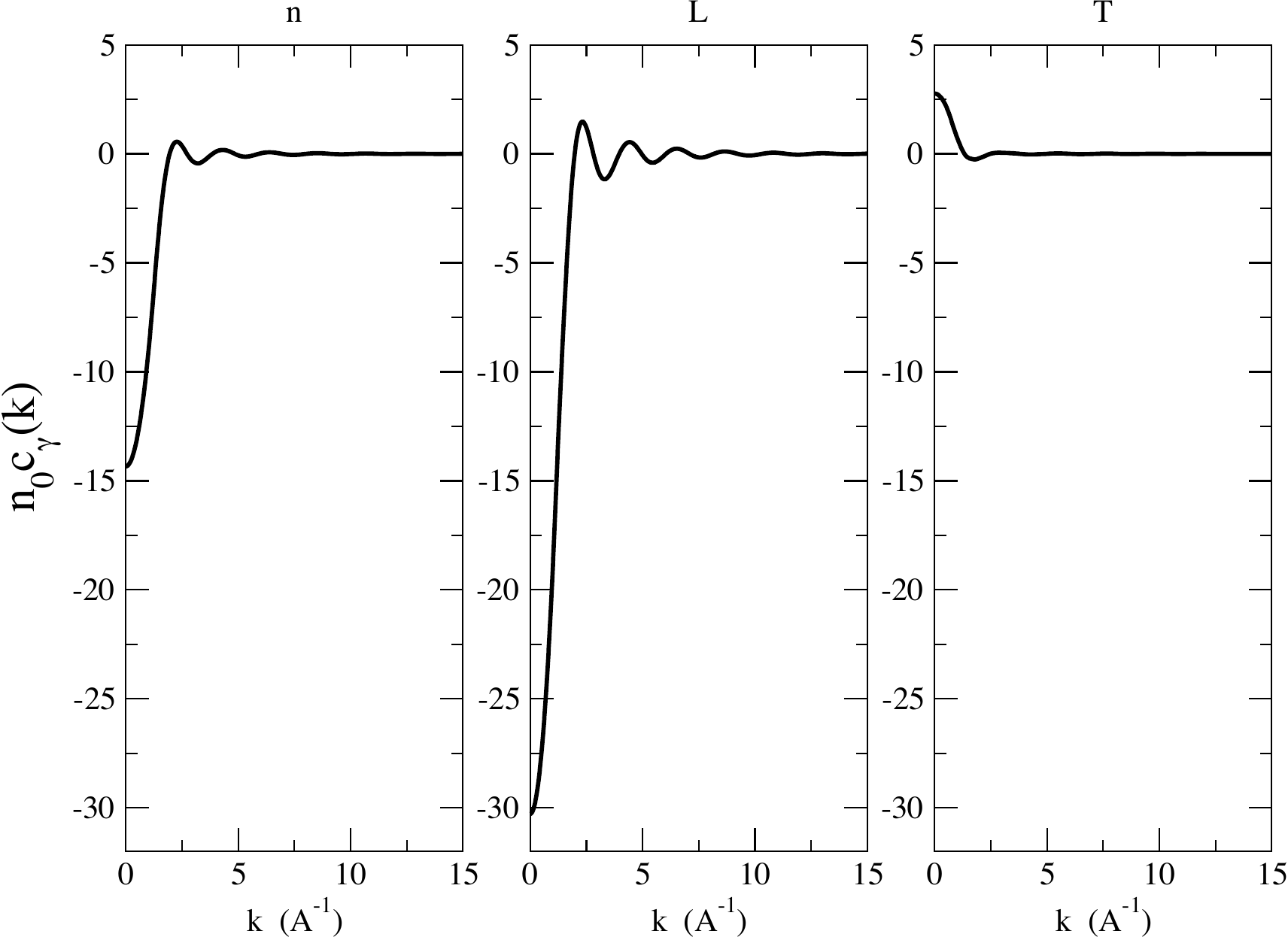}
	\caption{ \label{fig:c_nLT_stockmayer}
Direct correlation functions for the Stockmayer model described in the text, computed from the susceptibilities by inversion of the OZ equation,  in conformity with eq.~\ref{eq:inverse_chi}.}
\end{figure}

From this general functional at dipolar order, we now restrict ourselves to the functional of the simplest conceivable model of dipolar solvent, the Stockmayer model. It is characterized
by a single Lennard-Jones center with parameters $\sigma_s, \epsilon_s$ and a dipole
$\boldsymbol{\mu}_s = \mu \Om$, where $\Om$ is the unitary orientation vector of the molecule.
In passing, the parameters are selected to make the model  look like water (similar density,   $n_0=0.033$ particles/\AA$^3$, particle size, $\sigma_s = 3$ \AA, and molecular dipole, $p=1.85 D$) although not tasting quite as water (no hydrogen bond in the model!).\cite{ramirez02} The dielectric constant can be estimated to be $\epsilon \simeq 140$ instead of 80.
For such purely dipolar model,  the charge-coupling term in the functional is irrelevant ($\chi_{nL} \equiv 0$ in eqs~\ref{eq:inverse_chi}) and the density-polarization (or density-charge) couplings are absent in eqs~\ref{eq:Fexc_P_k-space_with_C}, \ref{eq:Fexc_P_k-space} and \ref{eq:Fexc_rho_k-space}. In Fig.~\ref{fig:chi_nLT_stockmayer} are displayed the susceptibilities 
$\chi_n(k), \chi_L(k), \chi_T(k)$ computed from the total correlation functions as in eqs~\ref{eq:chi_n_dipolar}. The latter functions were obtained by performing MD simulations of the pure solvent model at $300 \, K$.  
The corresponding direct correlation functions have characteristic forms that are plotted in Fig.~\ref{fig:c_nLT_stockmayer}.

A molecular solute embedded in the solvent will create an external potential which can be written as
\be
V_{ext}(\rr,\Om) = \Phi_{n}(\rr) - \mu \mathbf{E}_q(\rr) \cdot \Om
\ee
with
\bea
\Phi_{n}(\rr) & = &\sum_{j=1}^{M} 4\epsilon_{sj} \left[ \left(\frac{\sigma_{sj}}{|\rr - \RR_j|}\right)^{12} - \left(\frac{\sigma_{sj}}{|\rr - \RR_j|}\right)^{6} \right] \label {eq:Phi_n}\\
\mathbf{E}_q(\rr) &= &   - \nabla \Phi_q(\rr) \label {eq:E_c} \\
\Phi_q(\rr) & = & \frac{1}{4 \pi \epsilon_0 }  \sum_{j=1}^{M}\frac{Q_j}{|\rr - \RR_j|}. \label {eq:Phi_c}
\eea
The  solute is here described by atomic sites $j$, located at $\mathbf{R}_j$, with Lennard-Jones parameters $\sigma_{sj}, \varepsilon_{sj}$ (using Lorentz-Berthelot mixing rules with respect to solvent LJ parameters), and point charges $Q_j$. As for the excess functional, the external functional can thus be written  be written as a functional of $n(\rr)$ and $\Pol$. 
\bea
\F_{ext}[n(r), \PP] &=& \int d\rr \Phi_n(\rr) \, n(\rr)  - \frac{1}{4 \pi \epsilon_0 } \int d\rr \, \mathbf{E}_q(\rr) \cdot \PP \\
& = &\int d\rr \Phi_n(\rr) \, n(\rr)  + \frac{1}{4 \pi \epsilon_0 } \int d\rr \, \Phi_q(\rr) \rho_P(\rr)
\eea
Last but not least, this turns out  to be also the case for the ideal functional\cite{ramirez02,ramirez05}
\bea
\F_{id}[n,\PP]  &=& k_BT \,\int d\rr \,  n(\rr)
\ln(\frac{n(\rr)}{n_0})-n(\rr)+n_0   \\
 &+&
k_BT \int d\rr \, n(\rr) \left( \ln \left[\frac{\Lm(\Pon)}{\sinh(\Lm(\Pon))} \right ] 
+ \Pon \,\Lm(\Pon) \right). \nn
\label{eq:omega-ideal} 
\eea
In the second, polarization term,  $\Ld$ designates the Langevin function and $\Lm$ its inverse;
 $P(\rr)$ is
the modulus of the polarization vector $\Pol$. The expansion of those two terms yields at leading, quadratic order
\bea
\F_{id}^q[n,\PP] &= &\frac{k_BT}{2 n_0}\int d\rr  \Delta n(\rr)^2 + \frac{3k_BT}{2 \mu^2}  \int d\rr \, \frac{\Pol^2}{n(\rr)} \nn \\
&\simeq &\frac{k_BT}{2 n_0}\int d\kk \,  \Delta n(\kk) \, \Delta n(-\kk)  + \frac{3k_BT}{2\mu^2 n_0}  \int d\kk \, \PP(\kk) \cdot \PP(-\kk) 
\label{eq:quadratic_entropy}
\eea
In the second, polarization term one can recognize $\alpha_d = \mu^2/3k_BT$, the equivalent polarizability of a dipole $\mu$
at temperature $T$. One can thus interpret this term as either a rotational entropic term, or the polarization free-energy in a medium with local electric susceptibility
$\chi_e(\rr) = \alpha_d n_0$

The previous equations for dipolar fluids give some insight on how the functional can be interpreted and eventually generalized to multipolar rather than dipolar fluids. It is useful to  notice in particular that 
the excess  free-energy in eqs~\ref{eq:Fexc_P_k-space}-\ref{eq:Fexc_rho_k-space} can be written as 
\be
\F_{exc}[n,\PP] = \F^q[n,\PP] - \F_{id}^q[n,\PP]
\ee
where $\F^q[n,\PP]$ is an exact quadratic expansion of the free-energy functional around the homogenous density $n_0/4 \pi$; it involves by definition the inverse of susceptibilities.  
 The susceptibilities  are thus the measurable quantities to be injected in the theory. $\F_{id}^q[n,\PP]$ as it stands is the exact leading order in the expansion of the ideal free-energy, involving both a translational and rotational entropy. This is indeed a very classical view when juggling between inverse susceptibilities and direct correlation functions in simple fluids\cite{hansen} The inclusion of the rotational contributions is worth to be sorted out.

\begin{figure}
	\includegraphics[width=8cm]{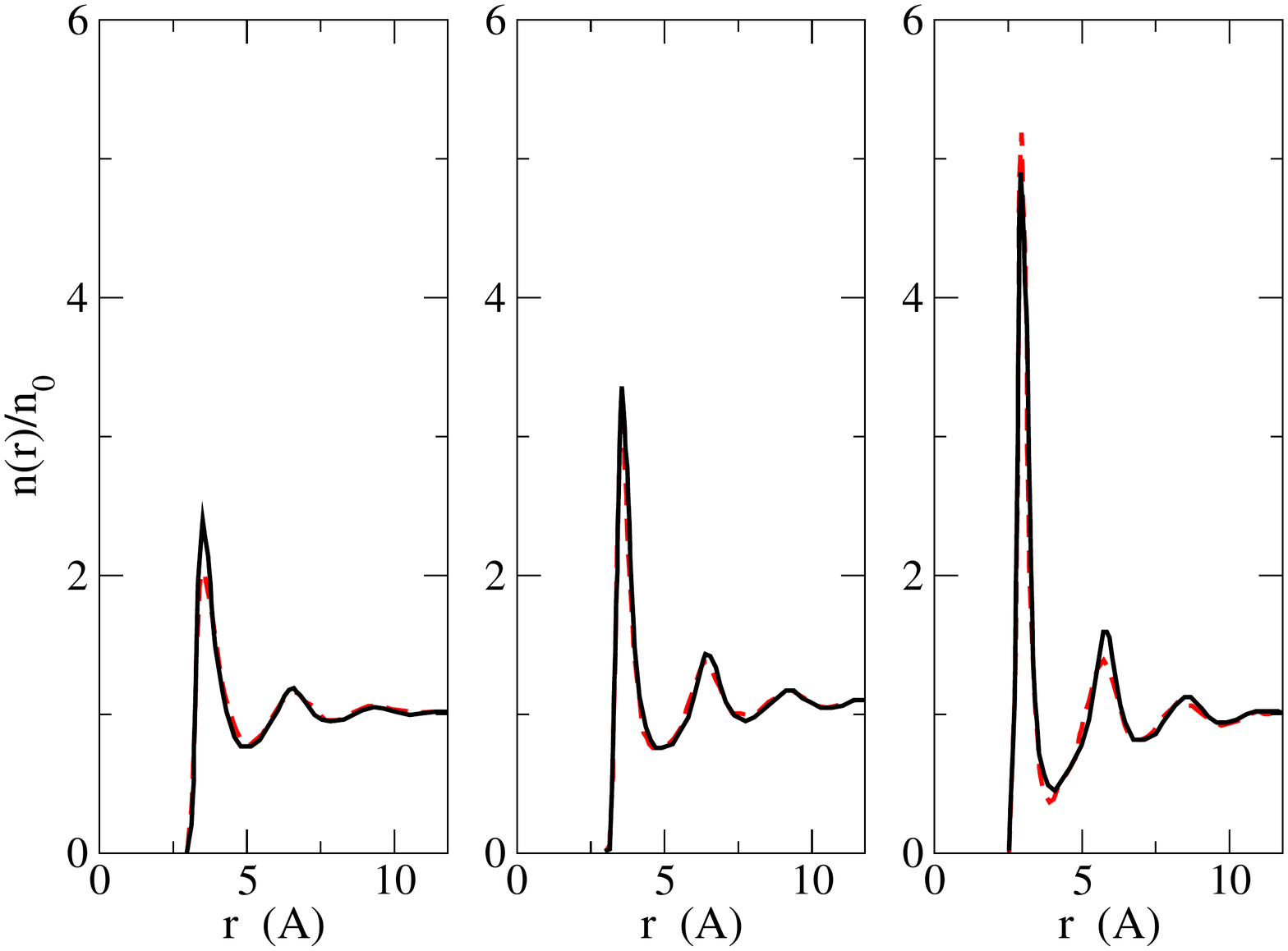}
	\includegraphics[width=8cm]{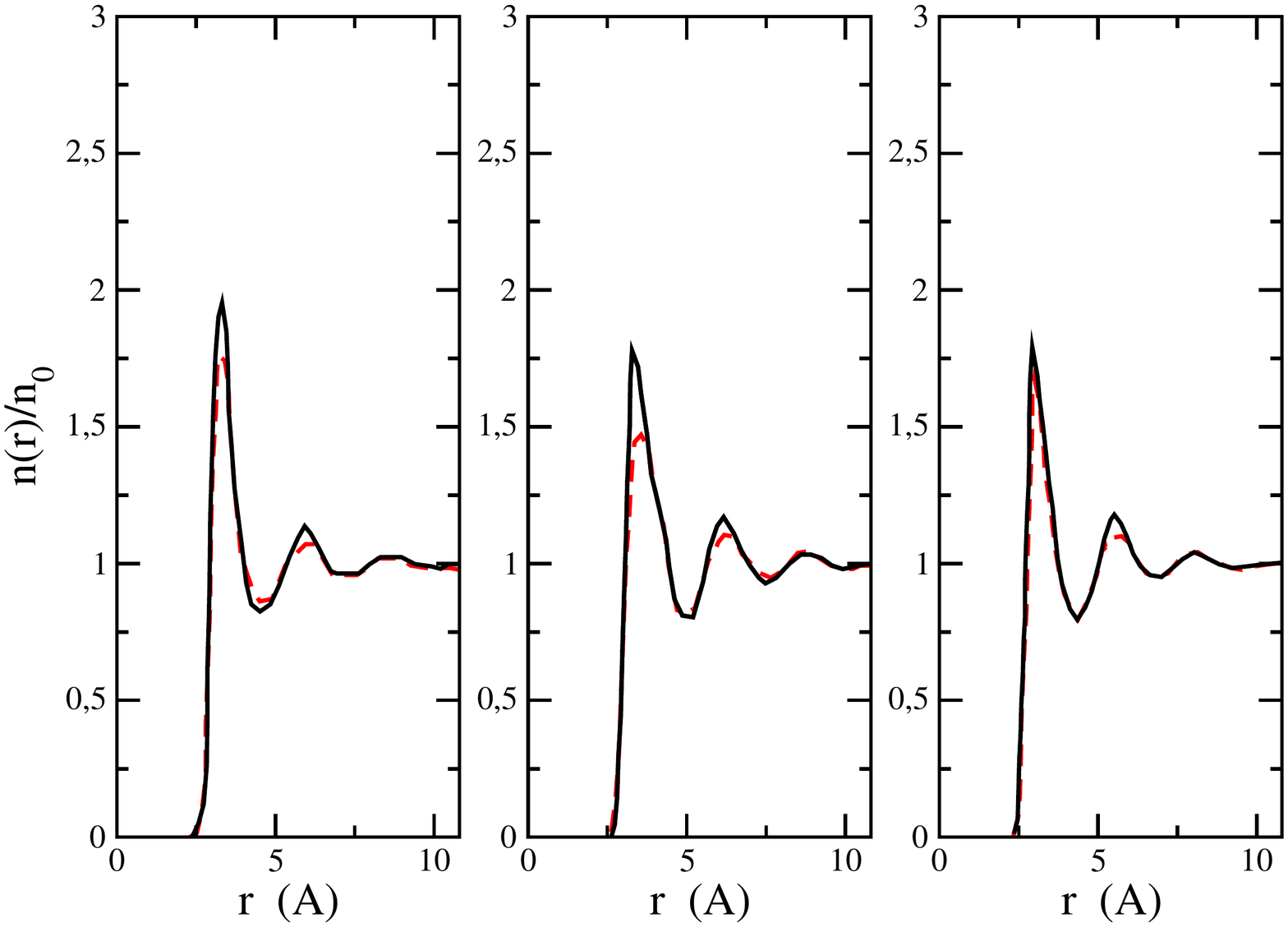}
	\caption{Top: Reduced density of the Stockmayer solvent around various solutes. MDFT results (solid black lines) are compared to MD simulation results (dashed red lines). From left to right: CH$_4$, Cl$^-$, K$^+$. Bottom: Same than for the various sites of an acetonitrile molecule dissolved in the Stockmayer solvent. From left to right: CH$_3$, C, N.
\label{fig:gr_CH4_Cl_K_CH3CN_stock}.}
\end{figure}

\begin{figure}
	\includegraphics[width=8cm]{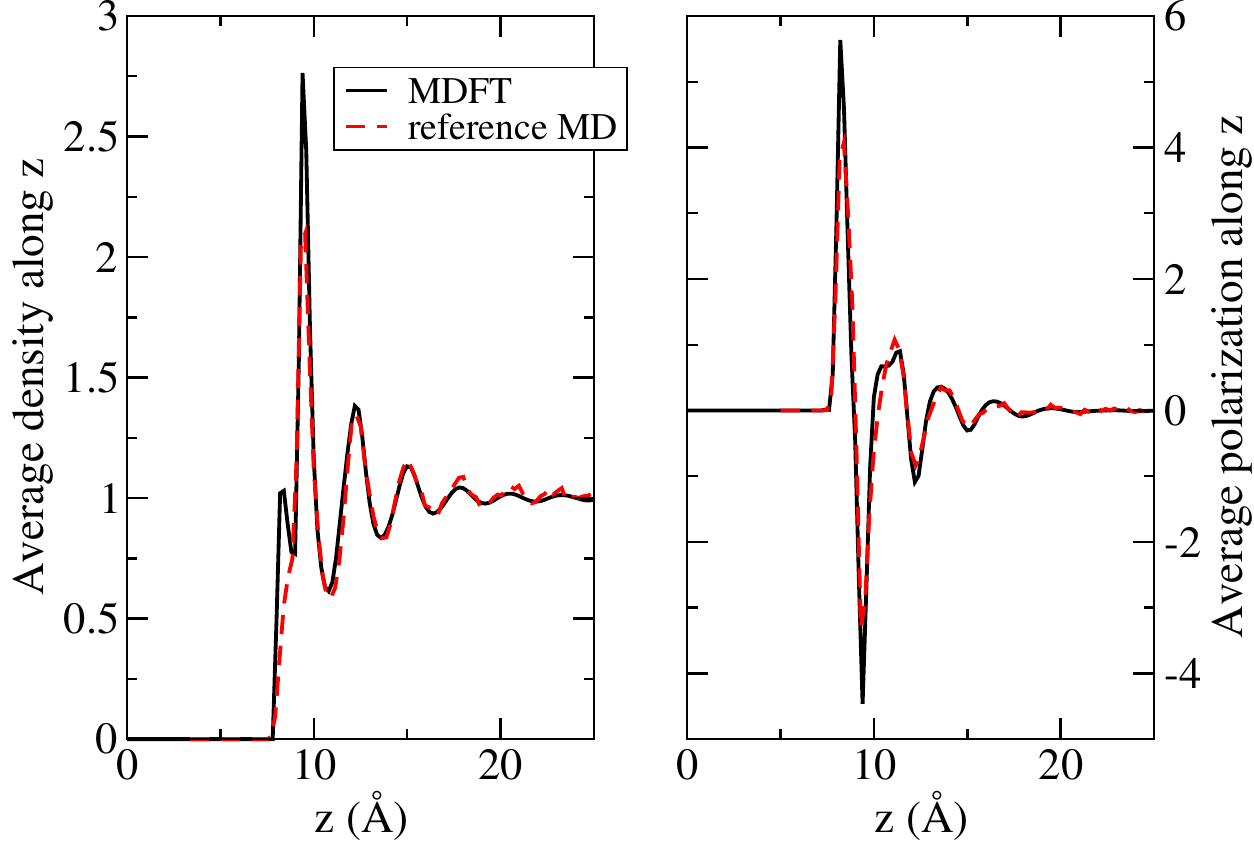}
	\caption{\label{fig:n_P_pyrophylite} Solvent number density and polarization density of a Stockmayer solvent as a function of the distance to an atomically-resolved clay surface (pyrophyllite). $z$ is the distance to the surface. MDFT results (black lines) are compared to reference all-atom molecular dynamic simulations (red lines);  Further details available in Ref.~\cite{levesque12_2} for details.}
\end{figure}

\begin{figure}
	\includegraphics[width=8cm]{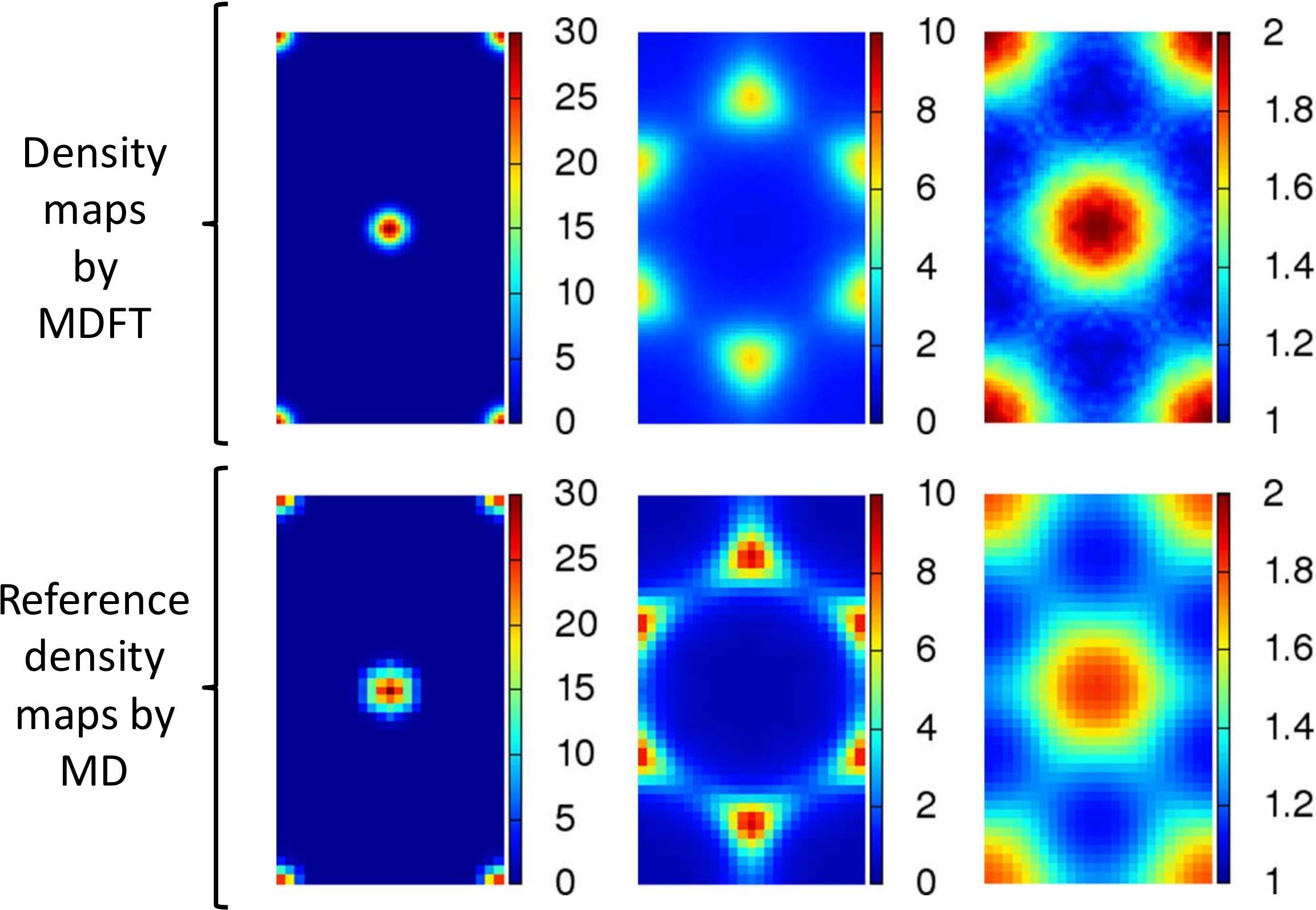}
	\caption{\label{fig:density-map_pyrophylite}Two-dimensional maps of the solvent number density
$n(\mathbf{r})/n_{0}$ in  three different planes close to a neutral clay surface, as calculated by molecular dynamics
(top) and HRF-MDFT (bottom). Those planes correspond to a prepeak  (left), the first maximum
(center) and second maximum (right) of the out-of plane mean solvent density.  See Ref.~\cite{levesque12_2} for details.}
\end{figure}

As a short illustration,  Fig.~\ref{fig:gr_CH4_Cl_K_CH3CN_stock}a shows the accuracy of the MDFT approach (within the HRF approximation, $\F_B[n,\PP] = 0$) for the microscopic structure of the Stockmayer solvent with properties described above around neutral and charged spherical solutes\cite{ramirez02,ramirez05,zhao11}. The MDFT results are compared to direct MD simulations of the solute embedded in the solvent. They do appear very satisfactory and account accurately for the shape of the peaks and their variation with charge and size (despite a slight overestimation of the first peak height for the neutral solute). Fig. ~\ref{fig:gr_CH4_Cl_K_CH3CN_stock}b illustrates the case of a multi-site polar molecule (here a three-site model of the acetonitrile molecule) with similar conclusions. An application to a more complex molecular system, namely the three-dimensional solvation structure close to an  atomically resolved surface of clay  is illustrated in Figs~\ref{fig:n_P_pyrophylite}-\ref{fig:density-map_pyrophylite} and  described further in Ref.~\cite{levesque12_2}. It is seen again that the agreement for the solvent structure, both in terms of  solvent density or solvent polarization density, is quite satisfactory. There is certainly room for improvement, in particular for the reproduction of the thermodynamic properties such as the solvation free-energies. This can be done by adding a spherical hard-sphere bridge which may seem  natural for such simple  solvent\cite{levesque12_1}, or learning from the exact bridge derived recently  for the Stockmayer liquid by Puibasset and Belloni\cite{belloni12}. We leave those possible refinements at the moment  and  shift immediately  to the description of more realistic solvent models.

\section{Generalization to multipolar order: The case of SPC-like water}

\subsection{Simple point charge model}
We start from SPC- or TIPnP-like representation of water, constituted by a single Lennard-Jones center, located on
the oxygen, and $m$ charges distributed on various sites. Each molecule is supposed to be rigid with position $\rr_i$ and
orientation $\omega_i$. For a given water configuration, we define the microscopic particle densities 
\bea
\label{eq:hatrhon}
  \hat{\rho}(\rr,\omega) &= & \sum^N_{i = 1} \delta(\rr-\rr_i) \delta(\omega - \omega_i), \\
  \hat{\rho}_n (\rr) & =& \sum^N_{i = 1} \delta(\rr-\rr_i)  = \int d\omega  \hat{\rho}(\rr,\omega),
\eea
and the charge and multipolar polarization density 
\bea
\label{eq:hatrhoc}
  \widehat{\rho}_c (\rr) & = & \sum^N_{i = 1} \sigma(\rr -\rr_i, \omega_i) =  \int d\rr' d\omega \, \sigma(\rr-\rr', \omega) \hat{\rho}(\rr', \omega), \\
    \widehat{\PP}_c (\rr) & = & \sum^N_{i = 1} \MU(\rr -\rr_i, \omega_i)  
     =  \int d\rr' d\omega \, \MU(\rr-\rr', \omega) \hat{\rho}(\rr', \omega).
\eea
$\sigma (\rr, \omega)$ is  the molecular charge density of a water molecule taken at the origin with orientation $\omega$ and,  according to the definition of  Refs~\cite{raineri93,bopp98}, $\MU(\rr, \omega)$ is the corresponding molecular polarization density: 
\bea
\sigrom & = & \sum_m q_m \, \delta\left(  \rr - \mathbf{s}_m(\omega)      \right),
\label{eq:sigma_water} \\
\MU(\rr,\omega) &= & \sum_m q_m \, \mathbf{s}_m(\omega)   \int_0^1 du \, \delta ( \rr - u \, \mathbf{s}_m(\omega) ),
\eea
where  $\mathbf{s}_m(\omega)$ indicates the location of the $m^{th}$ atomic site for a given $\omega$.   It can be easily checked that molecular charge and polarization densities are linked by the usual relation $\sigma(\rr,\omega) = -\mathbf{\nabla} \cdot \MU(\rr,\omega) $. In k-space
\bea
\MU(\kk,\omega) &= & -i \, \sum_m q_m \frac{\mathbf{s}_m(\omega) }{\kk \cdot \mathbf{s}_m(\omega) } \left( e^{i \, \kk \cdot \mathbf{s}_m(\omega) } -1 \right) \\
\label{eq:mukom}
&= & \MU(\omega) + \frac{i}{2} \sum_m q_m \left( \kk \cdot \mathbf{s}_m(\omega) \right)  \mathbf{s}_m(\omega) + ... ,
\eea
with $\MU(\omega) = \sum_m q_m \, \mathbf{s}_j(\omega) =  \mu \, \Om$  ($\mu$ the molecule dipole; for water, $\Om$ is
the unit vector along the O-H bonds angle bisector) The molecular polarization density thus reduces to a molecular dipole located at the origin at dominant order, but it does include  the complete multipole series when all other orders are considered. 

\subsection{A useful theorem}
These definitions being set, our statement starts from the observation that the Hamiltonian of  $N$ water molecules in the presence of an embedded solute, described by an external  molecular force field,  can be written as
 \be
\label{eq:H_rho_rhoc}
H_N = T + U + \int \dr \, \hat{\rho}_n (\rr) \Phi_n(\rr) - \int \dr \, \hat{\PP}_c (\rr) \cdot \EE_c(\rr),
\ee
where $T$ and $U$ are  the water kinetic and  pair-wise potential energy, respectively. $\Phi_n(\rr), \, \EE_c(\rr)$ denote the value of the
external Lennard-Jones potential and electric field at position $\mathbf{r}$,  as defined in eqs~\ref{eq:Phi_n}-\ref{eq:Phi_c}.

Following  the original derivation of Evans\cite{evans79} and extending  it to four independent external field variables, 
$\Phi_n(\rr)$ and $E_{x,y,z}(\rr)$, instead of just one, it can be easily proved  that the grand potential $\Omega$ of the solute-water system
at a given water chemical potential may be expressed as a functional of the one-particle number density, $n(\rr) =  \left< \hat{\rho}_n(\rr) \right>$,  and of the one-particle polarization density, $\PP_c(\rr) =  \left< \hat{\PP}_c (\rr) \right>$, that is $\Omega= \Omega[n(\rr),\PP_c(\rr)]$. Minimization of this functional with respect to those two fields yields the equilibrium densities and the value of the grand potential.  Taking as a reference the bulk water system at the same chemical potential, with number density $n_0$ and grand potential $\Theta_0$, the same properties hold true for the solvation free energy 
$\F[n(\rr),\PP(\rr)] = \Omega[n(\rr),\PP(\rr)] - \Omega_0$.  This constitutes our useful theorem. See Appendix \ref{a1} for a careful derivation.

The question is now how to infer this functional rather than the more general form involving  $\rhorom$ written in Sec.~\ref{sec:general_functional}.

\subsection{The multipolar functional}

To this end, we can simply refer to Sec. \ref{sec:dipolar} and replace the dipolar polarisation $\PP(\rr)$ by its multipolar equivalent $\PP_c(\rr)$. Although we know for sure that the whole functional 
could be expressed as functional of $n(\rr)$ and $\PP_c(\rr)$, the expression of the ideal term in this  way is very hard to infer, and we decide at this point to keep its full, exact expression in terms of 
$\rhorom$ (a sort of reminiscence of the Kohn-Sham idea in electronic structure when one has to resort to the orbitals instead of the density for the kinetic energy part). This leads to
\bea 
\F[\rho]& =& k_BT \int d\rr d\omega \left[ \rhorom){\rm ln}\left(\frac{8\pi^2 \rhorom}{n_0}\right )- \rhorom)+\frac{n_0}{8\pi^2} \right]   \label{eq:F_total_water} \\
& + & \int d\rr \Phi_n(\rr) \, n(\rr)  - \frac{1}{4 \pi \epsilon_0 } \int d\rr \, \mathbf{E}_q(\rr) \cdot \Pol - \F_{id}^q[\rho] \nn \\
&-& \frac{k_BT}{2 n_0}  \int d\kk \, \chi_n^{-1}(k) \, \Delta n(\kk) \, \Delta n(-\kk) + \frac{i  k_BT}{\mu_0 n_0}  \int d\kk \, \chi_{nL}^{-1}(k) \, P_{c,L}(\kk) \Delta n(-\kk) \nn \\
&+& \frac{1}{8\pi \epsilon_0} \left[ \int d\kk \, 4 \pi \chi_L^{-1}(k) \, \PP_{c,L}(\kk) \cdot \PP_{c,L}(-\kk) + 
\int d\kk \,  \chi_T^{-1}(k) \, \PP_{c,T}(\kk) \cdot \PP_{c,T}(-\kk) \right] \nn \\
&+& \F_B[n,\PP]   \nn
\eea
Two questions arise at this point: how to (i) generalize the susceptibilities to the model with distributed partial charges, and (ii) decide for an expression of the quadratic (linear response) entropy. We go to those questions in turn.

\begin{figure}
	\includegraphics[width=8cm]{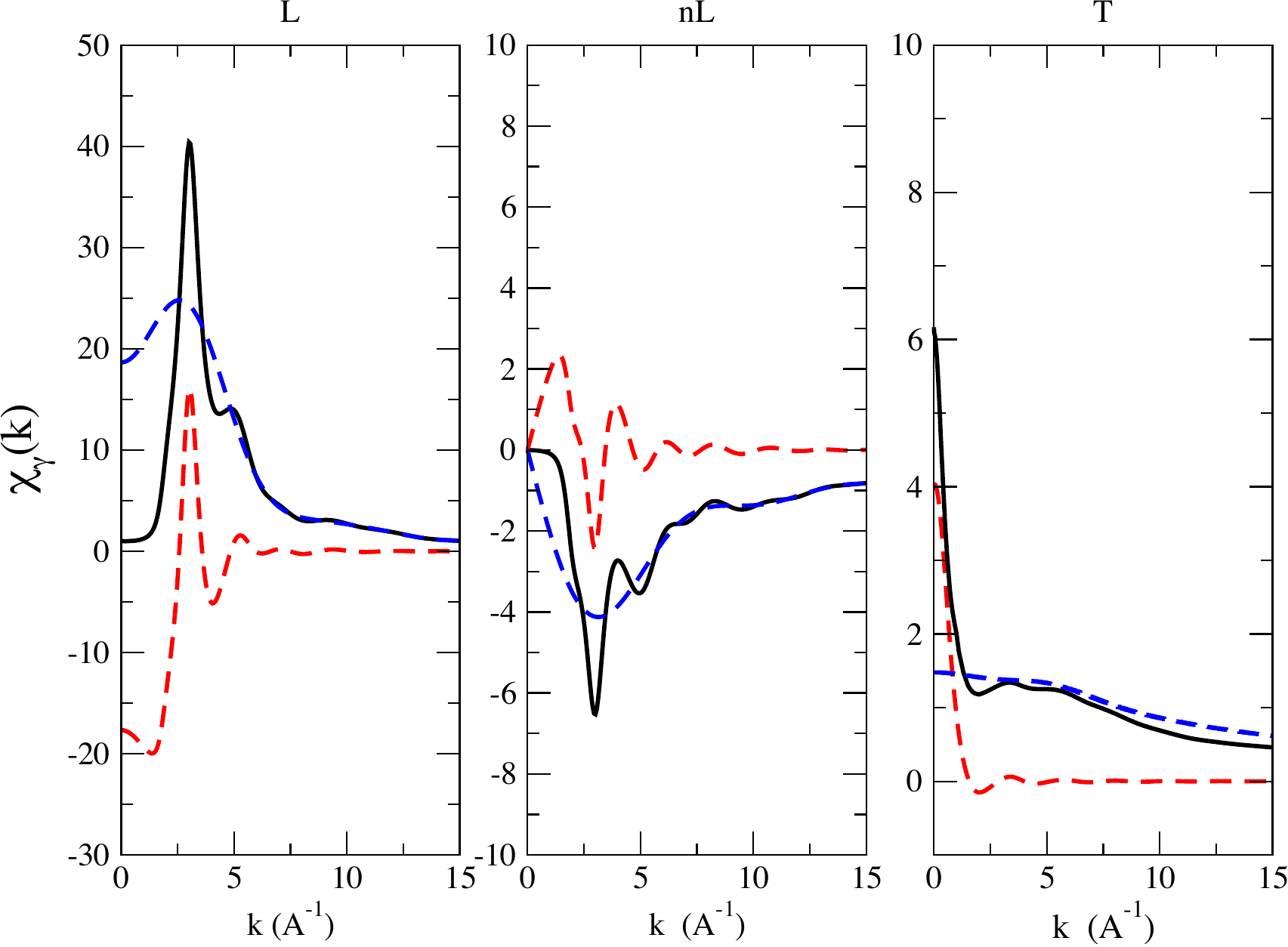}
	\caption{ \label{fig:chi_SPCE}
Susceptibilities $\chi_\gamma(k)$ with $\gamma = L, nL, T$ computed for  SPC/E water  by MD simulations (black solid lines). The self and distinct contributions are plotted in dashed lines (blue and red curves, respectively).}
\end{figure}

The point-charge-based, multipolar analog of the dipolar susceptibilities of \ref{sec:dipolar} can be defined as\cite{bopp96}
\bea 
\chi_L(k) &= & 1 - \frac{1}{\epsilon_L(k)} = \frac{S_c(k)}{\mu_0^2}  \\
\chi_{nL}(k) & = &  \frac{S_{nc}(k)}{\mu_0} \\
\chi_T(k)  &= & \frac{\epsilon_T -1}{4\pi} =  \frac{S_T(k)}{\mu_0^2}
\eea
with again an effective dipole defined by $\mu_0 = (\epsilon_0/\beta n_0)^{1/2}$, and
\bea
S_c(k)   &= &    \left< \hat{\PP}_{c,L}(\kk) \cdot \hat{\PP}_{c,L}(-\kk) \right> \\
& = & \left< \hat{\rho}_c(\kk) \hat{\rho}_c(-\kk) \right>/k^2 \\
S_{nc} &= & \left< \hat{\rho}_c(\kk) \Delta \hat{n}(-\kk) \right>/k \\
S_T(k) & = & \left< \hat{\PP}_{c,T}(\kk) \cdot \hat{\PP}_{c,T}(-\kk) \right> 
\eea
Applying the definitions above, these quantities can be decomposed into the sum of a self, intra-molecular part  (s) and distinct, inter-molecular part (d), which read in the case of 
SPC/E  water (3 sites with $q_H = +0.4238$ and $q_O = -2 q_H$) 
\bea
k^2 \, S_L^s(k) & = & 2 q_H^2 \left[ 3 + \frac{sin(k \, d_{HH})}{k \, d_{HH}} - 4 \frac{sin(k \, d_{OH})}{k \, d_{OH}} \right]  \\
k^2 \, S_L^d(k) & = & 4 n_0  q_H^2 \left( h_{OO}(k) + h_{HH}(k) - 2h_{OH}(k) \right)
\eea
for the charge-charge correlations,  
\bea
k \, S_{nL}^s(k) & = &  2 q_H \left[ \frac{sin(k \, d_{OH})}{k \, d_{OH}} - 1 \right] \\
k \, S_{nL}^d(k) & = &  2 n_0 q_H \left( h_{OH}(k) - h_{OO}(k) \right)
\eea
The  transverse inter-molecular correlation function cannot be recast in terms of the site-site pair distribution functions and has to be evaluated 
directly in k-space\cite{bopp98}.  We have computed all those various susceptibilities by molecular dynamics for SPC/E water; they are displayed in Fig.~\ref{fig:chi_SPCE}.
Knowing those functions,  one can define the inverse of the susceptibility matrix as in  eqs~\ref{eq:inverse_chi}. 

From there one can proceed in two ways. The first one is to feed directly the functional of eq.~\ref{eq:F_total_water}
with those inverse susceptibilities. One should then replace consistently the rotational
entropy term quadratic in $\Pol$ by a more general one involving the rotational conditional probability $\alpha(\rr,\Omega) = \rho(\rr,\Omega)/n(\rr)$; see appendix. This was the strategy adopted in Ref.~\cite{jeanmairet_molecular_2013}. The alternative, probably more consistent,  is 
to define the equivalent of the direct correlations in eqs~\ref{eq:inverse_chi} by extracting the intramolecular (ideal-gas) contributions in the inverse susceptibility matrix. This amounts to identify
 $\F_{id}^q[\rho] $  to the self interaction part in the quadratic expression of $\F[n,\PP]$,i.e.,
\bea
\F_{id}^q[\rho] &= &  \frac{k_BT}{2 n_0}  \int d\kk \, \chi_n^{s \, -1}(k) \, \Delta n(\kk) \, \Delta n(-\kk) + \frac{i  k_BT}{\mu_0 n_0}  \int d\kk \, \chi_{nL}^{s \,  -1}(k) \, P_{c,L}(\kk) \Delta n(-\kk) \nn \\
&+& \frac{1}{8\pi \epsilon_0}  \int d\kk \, 4 \pi \chi_L^{s \,  -1}(k) \, \PP_{c,L}(\kk) \cdot \PP_{c,L}(-\kk) + \frac{1}{8\pi \epsilon_0}   \nn 
\int d\kk \,  \chi_T^{s \, -1}(k) \, \PP_{c,T}(\kk) \cdot \PP_{c,T}(-\kk)
\eea
and thus to define the direct correlation functions by
\be
n_0 C_\gamma(k) =  \chi_\gamma^{s \, -1}(k) - \chi_\gamma^{-1}(k) 
\ee
for $\gamma = n, L, nL$, and $T$.

\begin{figure}
	\includegraphics[width=8cm]{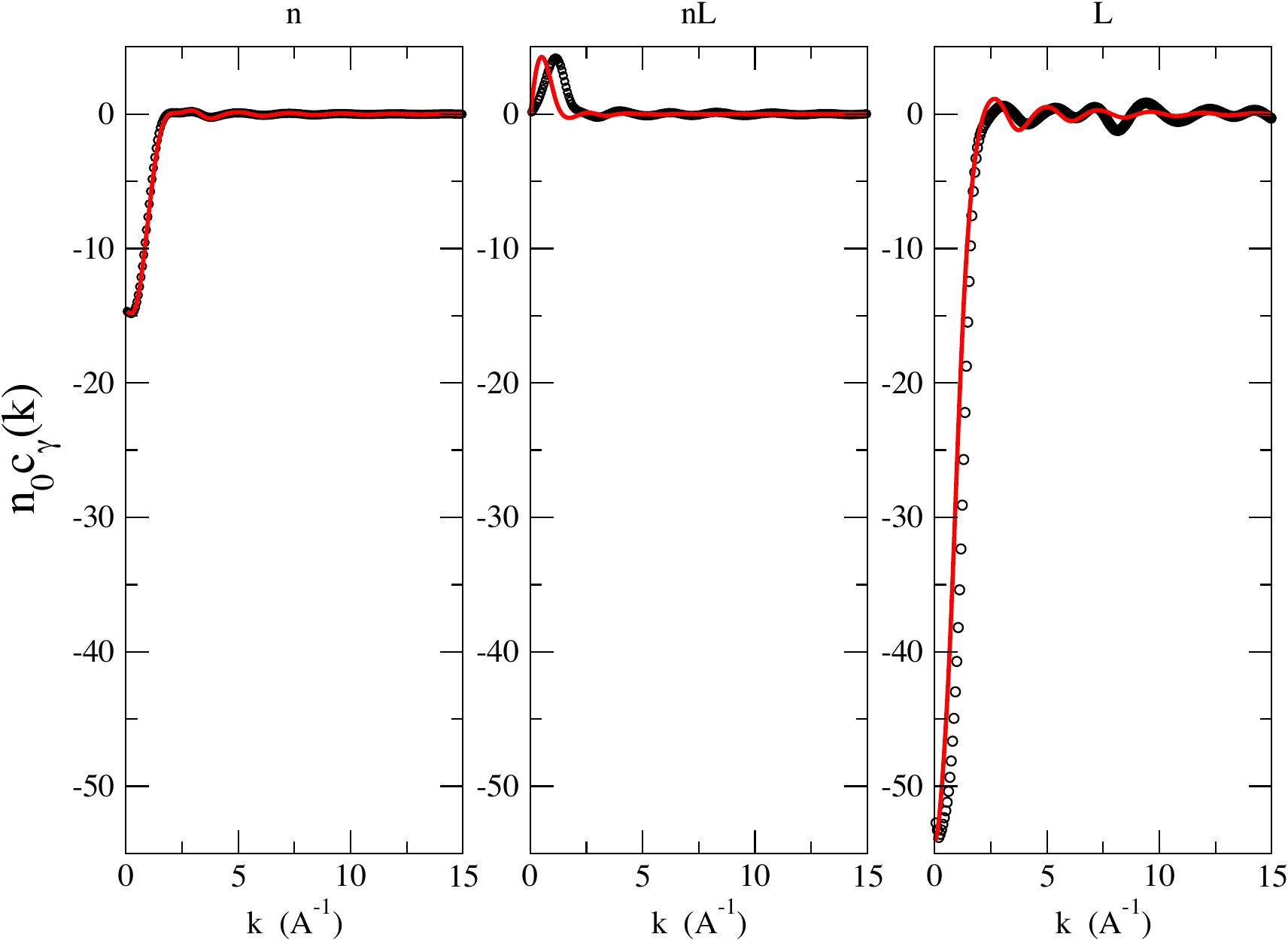}
	\caption{ \label{fig:c_n_nL_L_SPCE}
Direct correlation functions for SPC/E water in the dipolar or multipolar approximation. In the latter case we represent $C_n, \,  (\mu_0/\mu) C_{nL}$
and $(\mu_0/\mu)^2 C_{L}$. The agreement is equivalent for the transverse part, with a behavior similar to that in Fig.~\ref{fig:c_nLT_stockmayer}.}
\end{figure}

\begin{figure}
	\includegraphics[width=8cm]{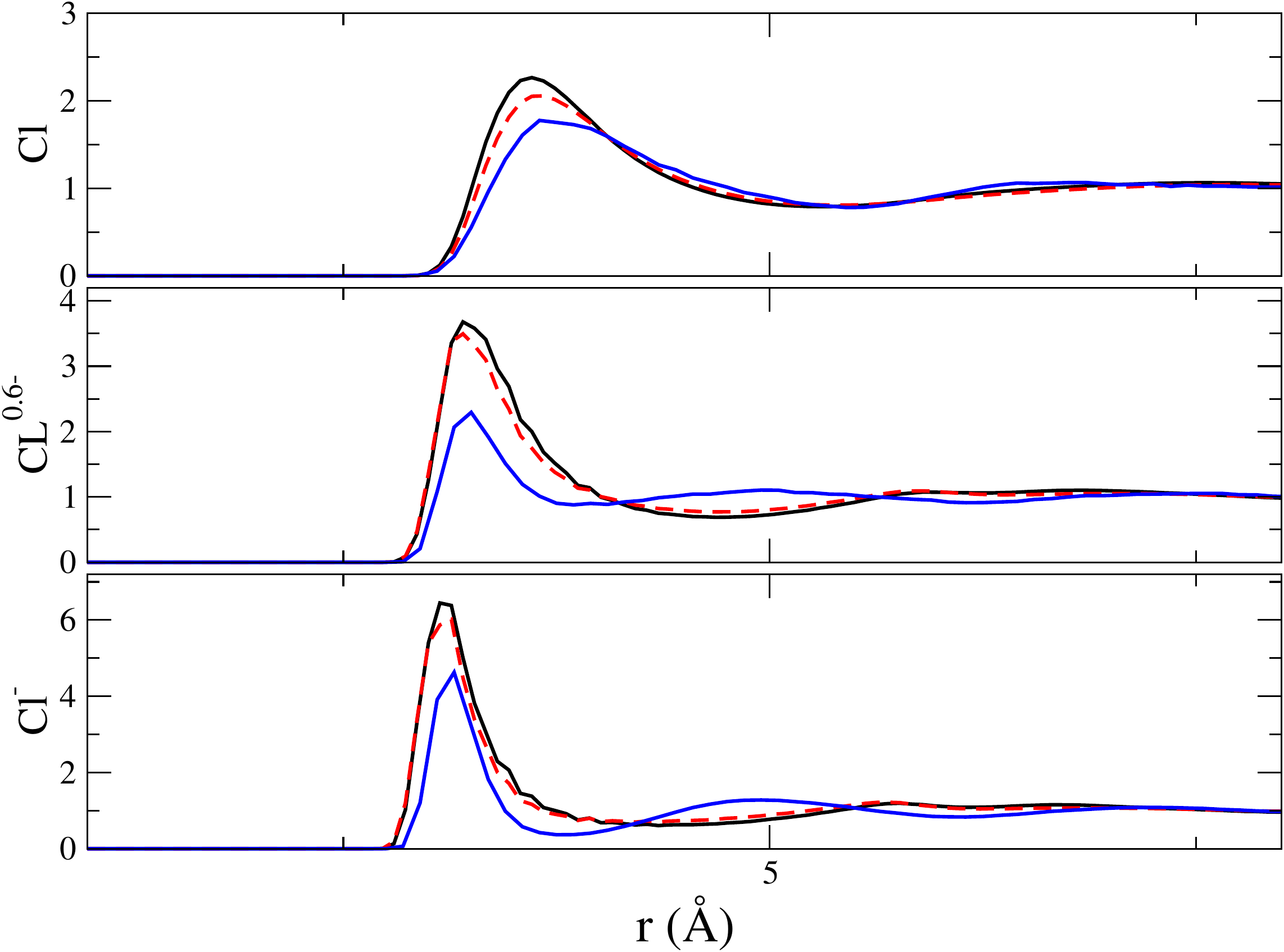}
	\caption{ \label{fig:gCl}
Radial distribution functions of, from top to bottom, neutral, $-\frac{3}{5}$ and -1 chloride. The MD simulations are in blue, while the results obtained by functional minimization with and without coupling between polarization and solvent density are in plain black and dashed red respectively.}
\end{figure}

\begin{figure}
	\includegraphics[width=8cm]{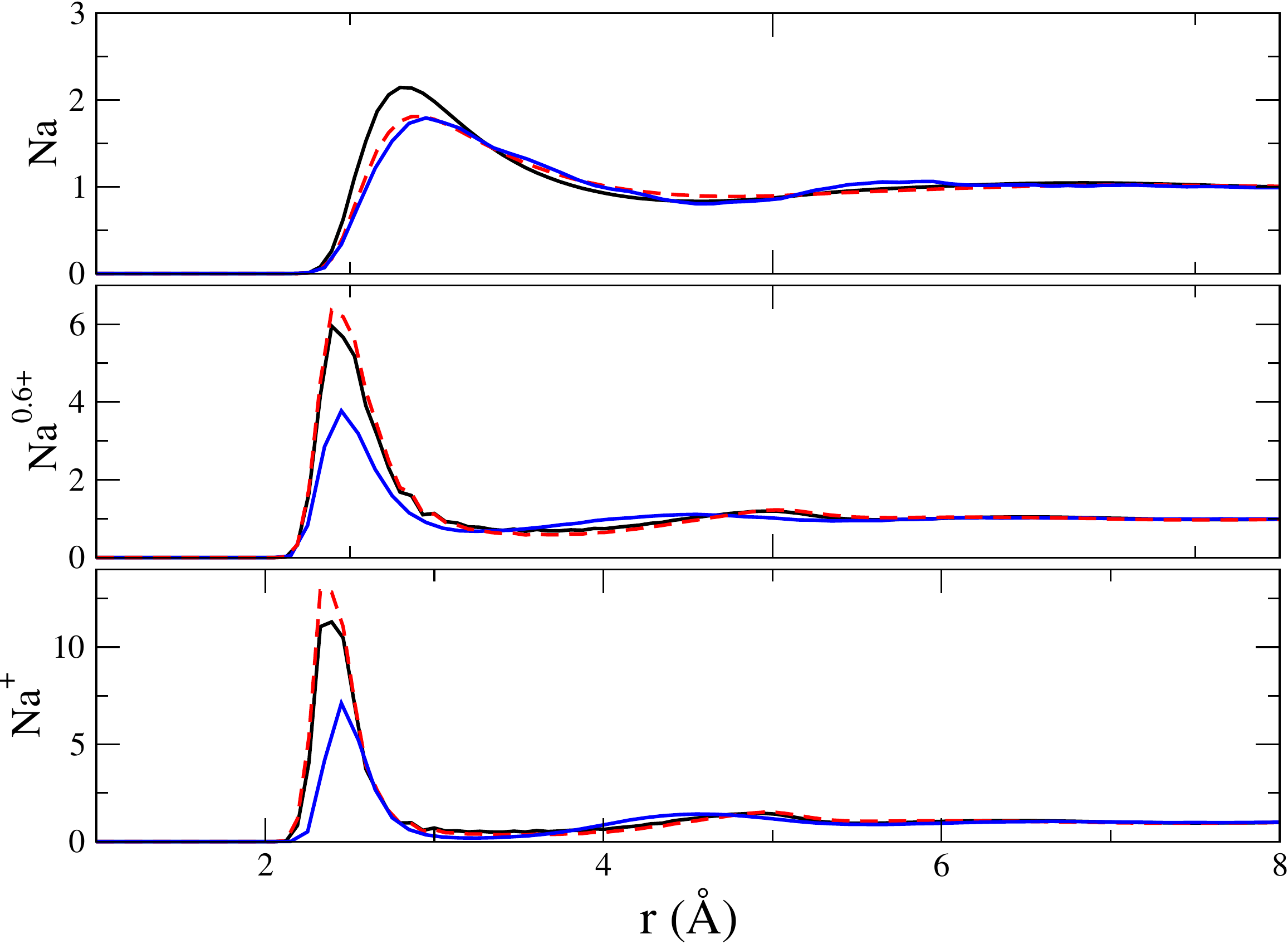}
	\caption{ \label{fig:gNa}
Radial distribution functions of, from top to bottom, neutral, $\frac{3}{5}$ and +1 sodium. The MD simulations are in blue, while the results obtained by functional minimization with and without coupling between polarization and solvent density are in plain black and dashed red respectively.}
\end{figure}

\begin{figure}
	\includegraphics[width=8cm]{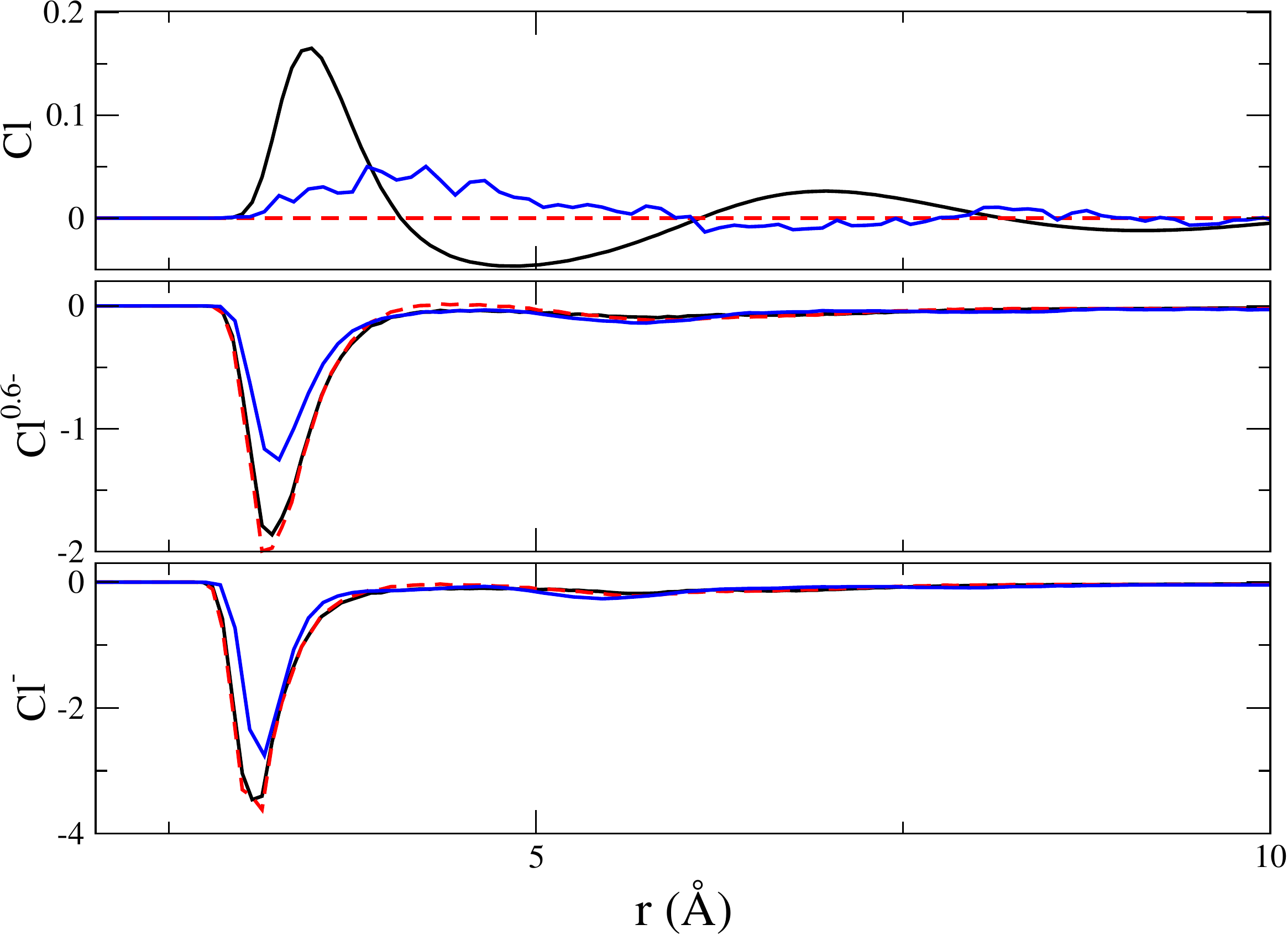}
	\caption{ \label{fig:polacl}
Radial polarization function ($\boldsymbol{P}(\boldsymbol{r})\cdot\boldsymbol{e}_r$) around neutral (top), $-\frac{3}{5}$ (middle) and -1 (bottom) chloride. The legend is the same than in Fig \ref{fig:gCl}.}
\end{figure}

\begin{figure}
	\includegraphics[width=8cm]{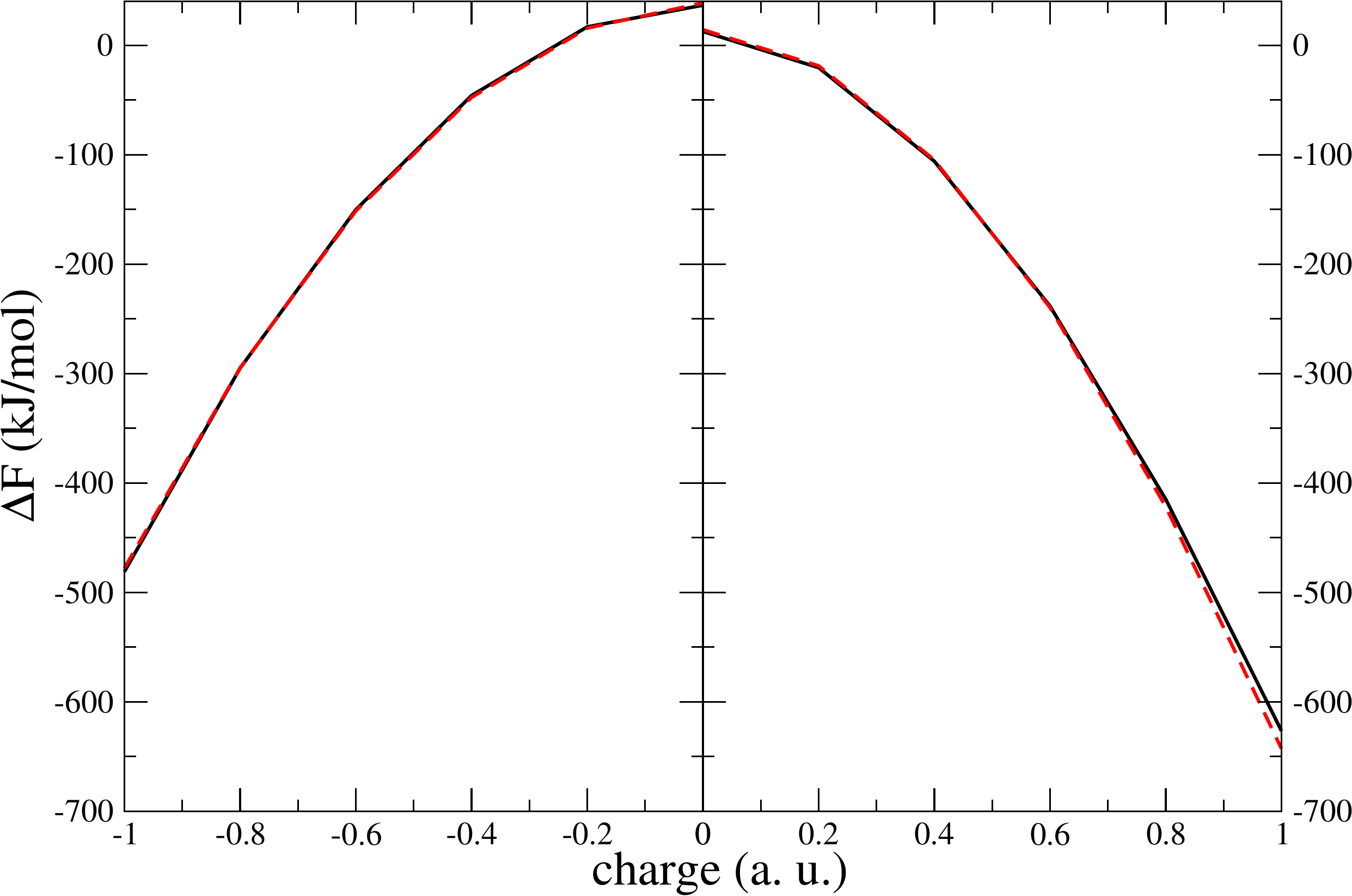}
	\caption{ \label{fig:FNaCl}
	Solvation free energy of chloride (left) with charge growing from -1 to 0, and sodium (right) with charge growing from 0 to +1. The black and dashed red lines have respectively been obtained by minimization of the functional with and without the coupling terms.}
\end{figure}

Injecting those definitions in the excess part of eq.~\ref{eq:F_total_water} leads to our final version of the  functional with an excess part defined by:
\bea
\F_{exc}[n,\PP] &=& - \frac{k_BT}{2}  \int d\kk \, C_{n}(k) \, \Delta n(\kk) \Delta n(-\kk)  - \frac{i \, k_BT}{\mu_0}  \int d\kk \, C_{nP}(k) \, P_{c,L}(\kk) \Delta n(-\kk)  \nn \\
& -& \frac{k_BT}{2 \mu_0^2}  \left[  \int d\kk \, C_L(k) \,  \PP_{c,L}(\kk) \cdot \PP_{c,L}(-\kk)  + \int d\kk \, C_T(k) \, \PP_{c,T}(\kk) \cdot \PP_{c,T}(-\kk)  \right] \nn \\
& + & \F_B[n,\PP]   \label{eq:Fexc_water_final} 
\eea
analog of eq.~\ref{eq:Fexc_P_k-space_with_C} with the correspondence 
\bea
C_n(k) &=& c_n(k) \nn \\
C_{nL}(k) & =& (\mu_0/\mu) \, c_{L,T}(k) \\
C_{L,T}& =& (\mu_0/\mu)^2 \, c_{L,T}(k) \nn
\eea
where $\mu$ is the water dipole. For the SPCE model, $\mu = 2.35 D$, and  $\mu_0/\mu = 0.134$. In Fig.~\ref{fig:c_n_nL_L_SPCE}, we compare the various direct correlation to their
dipolar analogs, computed as in Sec.~\ref{sec:dipolar}. Surprisingly, they appear very similar, with a slightly shifted peak for the density--polarization coupling term ($nL$). Oscillations appear above $\sim 7 ~$\AA for the longitudinal polarization term.
We now test this theory on archetypal solutes, that are Lennard-Jones balls with a partial charge at the center. The cationic solute is supposed to mimic sodium and have the following Lennard-Jones parameters ($\sigma=2.584$ \AA, $\epsilon=0.13$ kJ.mol$^{-1}$), while the anionic solute is a chloride model ($\sigma=4.035$ \AA, $\epsilon=0.51$ kJ.mol$^{-1}$). We first look at the effect of the proper treatment of the particle density--polarization coupling on the solvation free energies in Fig.\ref{fig:FNaCl}, where it is shown for "sodium" with a partial charge growing from 0 to +1 and for "chloride" for a charge decreasing from 0 to -1 obtained by functional minimization with and without the coupling term. In all case the results obtained with and without density--polarization coupling are within 3\%, even for ions as charged as +1.
The incapacity of the functional truncated at second order to properly describe the solvation structures of such charged solutes have already been be reported\cite{jeanmairet_molecular_2013}. In summary, because of its inherent lack of three-body and higher orders interactions like hydrogen-bonds, the first pick of the radial distribution function is too high and the second peak is too far.
In Fig.\ref{fig:gCl} and \ref{fig:gNa} we show the effect of the coupling and compare to reference molecular dynamics results. Once again, we stress out that the two functionals differ not only by the presence of the particle density polarization coupling term, but also by slightly different correlation functions. It is clear that the proper treatment of the particle density polarization coupling has very little effect on the radial distribution functions. It slightly corrects the height of the first peak for the positively charge sodium, but worsens the neutral sodium and chloride cases. The only qualitative effect of the correction introduced here is that is able to induce polarization for neutral, hydrophobic solutes. In Fig \ref{fig:polacl} we show that for charged solutes, density-polarization coupling has no or little effect. For neutral solutes, (say, a methane), one expects some polarization to be induced by a preferential orientation of the solvent around the repulsive pocket, for instance for entropic or purely geometric reasons. The former functional was not able to tackle such subtle effect: no electrostatic interactions with the solute implied no polarization of the solvent. We now have at least the good qualitative behavior.

\section{Conclusion}
In this article we included the density--polarization coupling to the molecular density functional theory for water. MDFT is now able to take into account the (multipolar) polarization that may arise from neutral hydrophobic solutes. Nevertheless, this density--polarization coupling show little effect on the solvation free energies and solvation structures.

With this generalization of the density+polarization theory for water, we reach here a point where it is clear that restraining the development of the functional to a quadratic order in density and polarization is not enough to capture all the properties of water, even if the polarization includes all multipoles of the solvent. To circumvent this problem it is necessary to introduce so-called bridge functionals or to go beyond our level of description of the direct correlation function. We proposed for instance an hard-sphere bridge and a three-body correction, that have the advantage of correcting a posteriori the pressure.\cite{jeanmairet3BPressure2015,volod_pressure_hnc_arxiv}

Another complementary route is to go beyond our level of description of the direct correlation function. The  basic equation derived in this work, eq. (48), is based on a quadratic expansion of the excess functional in the density and multipolar polarisation density; as such, when converted to a direct correlation function, it does include higher-order rotational invariants than just the dipolar ones described in Sec. 3 -- it does include \emph{all} those describing the multipole-multipole interaction. This is not equivalent, however, to expressing the functional at second order in $\rho\left( \boldsymbol{r},\omega \right)$, that is eq. (4), and using the full expansion of the angular-dependent direct correlation function defined as in eq. (5)-(6); such expansion contains in principle more rotational invariants and more information about the reference homogeneous fluid fluctuations than the $n$-$\boldsymbol{P}$ expansion does. The efficient implementation of eq. (4)-(6) using position and orientation fast (generalized) Fourier transforms, and the comparison to the more restricted approach described here, are currently underway.

\appendix

\section{The grand potential is a unique functional of the molecular density and polarization}\label{a1}

We first recall the expression of the classical trace, and the usual definitions for the functionals of the probability density $f$ of a system of N particles\cite{evans79}:

\begin{equation}
\text{Tr}_{\text{cl}}=\sum_{N=0}^\infty\frac{1}{h^{3N}N!}\int \text{d}\boldsymbol{r}_1\ldots \text{d}\boldsymbol{r}_N \int \text{d}\boldsymbol{p}_1 \ldots \text{d}\boldsymbol{p}_1
\end{equation}
where $\boldsymbol{r}_i$ and $\boldsymbol{p}_i$ stand for the position and momentum of particle $i$.
$H_N$ is the hamiltonian for $N$ particles, each of chemical potential $\mu$, at temperature $T$, with $\beta=(k_BT)^{-1}$.
\begin{equation}
\Omega\left[f\right]=\mathrm{Tr_{cl}}\left[f(H_{N}-\mu N+\beta^{-1}\ln f)\right],\label{eq:DefThetaTRace}
\end{equation}
\begin{equation}
\Omega\left[f_{0}\right]=-\mathrm{k_{B}}T\ln\Xi=\Omega,\label{eq:Theta0=00003DTheta(f0)}
\end{equation}
\begin{equation}
\Omega\left[f\right]>\Omega\left[f_{0}\right],\text{ if  \ensuremath{f\neq f_{0}}}.\label{eq:Thetaf>theta}
\end{equation}
Next, we introduce two functionals:
\begin{equation}
{\cal F}[n(\boldsymbol{r}),\boldsymbol{P}(\boldsymbol{r})]=\mathrm{Tr_{cl}}\left[f_{0}\left(T+U+\beta^{-1}\ln f_{0}\right)\right],\label{eq:F(n,P)}
\end{equation}
and
\begin{equation}
\Omega_{\mathrm{{\cal V}}}[n(\boldsymbol{r}),\boldsymbol{P}(\boldsymbol{r})]={\cal F}[n(\boldsymbol{r}),\boldsymbol{P}(\boldsymbol{r})]+\int\Psi_{n}(\boldsymbol{r})n(\boldsymbol{r})\mathrm{d}\boldsymbol{r}-\int\boldsymbol{P}(\boldsymbol{r})\cdot\boldsymbol{E}(\boldsymbol{r})\mathrm{d}\boldsymbol{r},\label{eq:Theta_V}
\end{equation}
where $\Psi_{n}(\boldsymbol{r})=\Phi_{n}(\boldsymbol{r})-\mu$. $\mathrm{{\cal V}}$
in Eq.~(\ref{eq:Theta_V}) is intended to remind that the functional depends
upon the form of the external potential. It is thus a compact notation
for $(\phi_{n},\boldsymbol{E})$.

The equilibrium probability density is a function of the external fields
$\Phi_{n}$ and $\boldsymbol{E}$. Since
\begin{equation}
n_{0}(\boldsymbol{r})=\mathrm{Tr_{cl}}\left[f_{0}\hat{n}(\boldsymbol{r})\right]
\end{equation}
and
\begin{equation}
\boldsymbol{P}_{0}(\boldsymbol{r})=\mathrm{Tr_{cl}}\left[f_{0}\hat{P}(\boldsymbol{r})\right],
\end{equation}
$n_{0}(\boldsymbol{r})$ and $\boldsymbol{P}_{0}(\boldsymbol{r})$ at equilibrium are also
functionals of these external fields
functionals. If we assume that it exist two tuples of different external
fields that induce the same equilibrium density and polarization $(n(\boldsymbol{r}),\boldsymbol{P}(\boldsymbol{r}))$.
At Hamiltonian $H_{N}$ we can associate equilibrium probability distribution
$f_{0}$ and grand potential $\Omega$. At Hamiltonian $H_{N}^{\prime}$
we associate $f_{0}^{\prime}$ and $\Omega^{\prime}$. Injecting in Eq. \ref{eq:Thetaf>theta}:
\begin{equation}
\Omega^{\prime}=\mathrm{Tr_{cl}}f_{0}^{\prime}\Bigl[\left(H_{N}^{\prime}-\mu N+\beta^{-1}\ln f_{0}^{\prime}\right)\Bigr]<\Omega\left[f_{0}\right],
\end{equation}
\begin{equation}
\Omega^{\prime}=\mathrm{Tr_{cl}}\Bigl[f_{0}^{\prime}\left(H_{N}^{\prime}-\mu N+\beta^{-1}\ln f_{0}^{\prime}\right)\Bigr]<\mathrm{Tr_{cl}}\Bigl[f_{0}\left(H_{N}^{\prime}-\mu N+\beta^{-1}\ln f_{0}\right)\Bigr],
\end{equation}
\begin{equation}
\Omega^{\prime}<\Omega+\int n_{0}(\boldsymbol{r})(\phi_{N}^{\prime}(\boldsymbol{r})-\phi_{N}(\boldsymbol{r}))\mathrm{d}\boldsymbol{r}-\int\boldsymbol{P}_{0}(\boldsymbol{r})\cdot(\boldsymbol{E}^{\prime}(\boldsymbol{r})-\boldsymbol{E}(\boldsymbol{r}))\mathrm{d}\boldsymbol{r}.\label{eq:theta<thetaprime+...}
\end{equation}
We exchange roles of $\Omega$ and $\Omega^{\prime}$, and thus
\begin{equation}
\Omega<\Omega^{\prime}+\int n_{0}(\boldsymbol{r})(\phi_{N}(\boldsymbol{r})-\phi_{N}^{\prime}(\boldsymbol{r}))\mathrm{d}\boldsymbol{r}-\int\boldsymbol{P}_{0}(\boldsymbol{r})\cdot(\boldsymbol{E}(\boldsymbol{r})-\boldsymbol{E}^{\prime}(\boldsymbol{r}))\mathrm{d}\boldsymbol{r}.\label{eq:thetaprime<tehta+...}
\end{equation}
By summing Eq.\ref{eq:theta<thetaprime+...} and \ref{eq:thetaprime<tehta+...} we obtain the absurd following inequality: $\Omega+\Omega^{\prime}<\Omega^{\prime}+\Omega$. As a consequence, to one tuple of density and polarization is associated
one and only one external potential. It follows that the probability
distribution is a unique functional of the equilibrium densities and that the functional given in Eq. \ref{eq:F(n,P)} is thus a unique functional of $n$
and $\boldsymbol{P}$. It is said universal in that its expression is the same
for all external potentials.

Furthermore, for equilibrium densities $n_{0}(\boldsymbol{r})$ and $\boldsymbol{P}_{0}(\boldsymbol{r})$,
this functional is equal to the grand potential:
\begin{equation}
\Omega_{{\cal V}}[n_{0}(\boldsymbol{r}),\boldsymbol{P}(\boldsymbol{r})]=\Omega.
\end{equation}
We now suppose there exists two molecular densities and polarizations, $n^{\prime}(\boldsymbol{r})$
and $\boldsymbol{P}^{\prime}(\boldsymbol{r})$, that are associated with the same probability distribution
$f^{\prime}$.
\begin{eqnarray*}
\Omega\left[f^{\prime}\right] & = & \mathrm{Tr_{cl}}\left[f^{\prime}(H_{N}-\mu N+\beta^{-1}\ln f^{\prime})\right]\\
 & = & {\cal F}\left[n^{\prime}(\boldsymbol{r}),\boldsymbol{P}^{\prime}(\boldsymbol{r})\right]+\int\Psi_{n}(\boldsymbol{r})n^{\prime}(\boldsymbol{r})\mathrm{d}\boldsymbol{r}-\int\boldsymbol{P}^{\prime}(\boldsymbol{r})\boldsymbol{E}(\boldsymbol{r})\mathrm{d}\boldsymbol{r}\\
 & = & \Omega_{{\cal V}}\left[n^{\prime}(\boldsymbol{r}),\boldsymbol{P}^{\prime}(\boldsymbol{r})\right].
\end{eqnarray*}
Thanks to Eq.~ \ref{eq:Thetaf>theta}, this leads to
\begin{equation}
\Omega_{{\cal V}}\left[n^{\prime}(\boldsymbol{r}),\boldsymbol{P}^{\prime}(\boldsymbol{r})\right]>\Omega_{{\cal V}}\left[n_{0}(\boldsymbol{r}),\boldsymbol{P}_{0}(\boldsymbol{r})\right].
\end{equation}
Thus, equilibrium densities $n_{0}(\boldsymbol{r})$ and $\boldsymbol{P}_{0}(\boldsymbol{r})$
minimize the functional $\Omega_{\cal V}\left[n(\boldsymbol{r}),\boldsymbol{P}(\boldsymbol{r})\right]$:

\begin{equation}
\left.\frac{\delta\Omega_{{\cal V}}\left[n(\boldsymbol{r}),\boldsymbol{P}(\boldsymbol{r})\right]}{\delta n(\boldsymbol{r})}\right|_{n_{0}}=\left.\frac{\delta\Omega_{{\cal V}}\left[n(\boldsymbol{r}),\boldsymbol{P}(\boldsymbol{r})\right]}{\delta\boldsymbol{P}(\boldsymbol{r})}\right|_{\boldsymbol{P}_{0}}=0.
\end{equation}

\section{Translational and rotational entropy}

We define  the conditional probability  $\alpharom$  of having a solvent orientation  $\omega$ at position $\rr$.
\be
\rhorom = \rhon \, \alpharom
\ee
 Thus
\be
\int d\Omega \, \alpharom = 1
\ee
and the homogeneous probability is $\alpha_0 = 1/8 \pi^2$.

The total entropy (ideal term) 
\be
\F_{id}  = k_BT \int d\rr d\omega \left[ \rhorom){\rm ln}\left(\frac{8\pi^2 \rhorom}{n_0}\right )
 - \rhorom)+\frac{n_0}{8\pi^2} \right] 
\ee
can be easily decomposed into a translational and rotational
contribution
\bea
\F_{id} & = & \F_{id}^{T} + \F_{id}^{R} \nn \\
\F_{id}^{T}[\rhon] & = & k_BT \int d\rr \left[ \rhon ln \left(\frac{\rhon}{n_0} \right)   - \rhon + n_0 \right]  \\
\F_{id}^{R}[\rhon,\alpharom] &= &k_BT \int d\rr  \, \rhon \int d\omega \left[ \alpharom \ln \left(\frac{\alpharom}{\alpha_0} \right)  - \alpharom + \alpha_0 \right] \nn
\eea
If the translational part is expanded at dominant order in $\Delta n(\rr) = n(\rr) - n_0$ and  the  rotational part at dominant order  in $\delta \alpharom = \alpharom - \alpha_0$, one gets
\bea
\F_{id}^{T}[\rhon] & = & \frac{k_BT}{2 n_0} \int d\rr \, \Delta n(\rr)^2 \\
\F_{id}^{R}[\rhon,\alpharom] &= & k_BT \int d\rr  \, \rhon \int d\omega \, \frac{(\alpharom - \alpha_0)^2}{2 \alpha_0}
\eea
The sum of those to terms gives the expression of $\F_{id}^{q}[\rho]$ to be injected in eq.~\ref{eq:F_total_water} (route number 1).

Further expanding $\alpharom$ into generalized spherical harmonics\cite{gray-gubbins-vol1}
\be
\alpharom = \alpha_0 + \sum_{m \ge 1} \sum_{\nu, \nu' = -m}^{m} \alpha^{m}_{\nu \nu'} (\rr) \, D^{m}_{\nu \nu'}(\omega)
\ee
then
\be
\F_{id}^{R}[\rhon,\alpharom] = \frac{k_BT}{2 \alpha_0}  \int d\rr \, \rhon \, \sum_{m \ge 1,\nu, \nu'} | \alpha^m_{\nu \nu'}(\rr)|^2.
\ee
Limiting the expansion to linear symmetry ($\nu' = 0$, orientation vector $\Om = (\sin \theta \cos \phi, \sin \theta \sin \phi, \cos \theta)$ instead of $\omega$, and $\alpha(\rr,\Om) = 2 \pi \rhorom $), one gets  at lowest order $m \le 1$, 
\be
\frac{1}{\alpha_0} \, \sum_{\nu=-1}^{1}  |\alpha_{\nu0}^{1}(\rr)|^2 = 3 \bar{\Om}(\rr)^2 
\ee
where $\bar{\Om}(\rr) = \int d\Om \, \Om \,  \alpha(\rr,\Om)$  defines the average orientation.
Finally, with the definition $\Pol = \mu \, n(\rr) \, \bar{\Om}(\rr)$, one gets 
\be
\F_{id}^{R}[\rhon,\alpharom]  = \frac{3 k_BT}{2\mu^2} \int d\rr \, \frac{\Pol^2}{n(\rr)}
\label{eq:FidRot_lin}
\ee
which is the expression quoted in the text as the quadratic rotational entropy, and can be recovered by expanding the general expression of eq.~\ref{eq:omega-ideal} at dominant order.\\

\begin{acknowledgments}
ML and DB are grateful to Luc Belloni for illuminating discussions and useful comparisons for the direct correlation functions of water.
\end{acknowledgments}

\bibliography{dft_08-02-2015,mdfteam} 
\bibliographystyle{unsrt}

\newpage

\end{document}